\documentclass[twocolumn]{aastex631}
\usepackage[utf8]{inputenc}
\usepackage{amsmath,amssymb}

\usepackage{graphicx}
\usepackage[caption=false]{subfig}
\usepackage{hyperref}
\usepackage{color}
\usepackage{nameref}
\usepackage{wrapfig}

\definecolor{asparagus}{rgb}{0.53, 0.66, 0.42}

\submitjournal{ApJS}

\shorttitle{RelSIM}
\shortauthors{F.\ Bacchini}

\newcommand\bb[1]{\mbox{\boldmath{$#1$}}}
\newcommand\grad{\bb{\nabla}}
\newcommand\bcdot{\,\bb{\cdot}\,}
\newcommand\bdbldot{\,\bb{:}\,}
\newcommand\btimes{\,\bb{\times}\,}

\newcommand{\omp}{\omega_\mathrm{p}}
\newcommand{\ompe}{\omega_{\mathrm{p},e}}
\newcommand{\ompi}{\omega_{\mathrm{p},i}}
\newcommand{\vecE}{\bb{E}}
\newcommand{\vecvbar}{\bb{\bar{v}}}
\newcommand{\vecubar}{\bb{\bar{u}}}
\newcommand{\vecB}{\bb{B}}
\newcommand{\vecJ}{\bb{J}}
\newcommand{\veceps}{\bb{\epsilon}}
\newcommand{\vecbeta}{\bb{\beta}}
\newcommand{\bgam}{\bar{\gamma}}

\newcommand{\vecv}{\bb{v}}
\newcommand{\vecu}{\bb{u}}
\newcommand{\vecx}{\bb{x}}

\newcommand{\pd}{\partial}
\newcommand{\rmd}{\mathrm{d}}
\newcommand{\be}{\begin{equation}}
\newcommand{\ee}{\end{equation}}
\newcommand{\ba}{\begin{aligned}}
\newcommand{\ea}{\end{aligned}}
\DeclareMathAlphabet\mathbfcal{OMS}{cmsy}{b}{n}
\begin{document}

\title{RelSIM: A Relativistic Semi-implicit Method for Particle-in-Cell Simulations}

\correspondingauthor{F.\ Bacchini}
\email{fabio.bacchini@kuleuven.be}

\author[0000-0002-7526-8154]{Fabio Bacchini}
\affiliation{Centre for mathematical Plasma Astrophysics, Department of Mathematics, KU Leuven, Celestijnenlaan 200B, B-3001 Leuven, Belgium}
\affiliation{Royal Belgian Institute for Space Aeronomy, Solar-Terrestrial Centre of Excellence, Ringlaan 3, 1180 Uccle, Belgium}

\begin{abstract}
We present a novel Relativistic Semi-Implicit Method (RelSIM) for particle-in-cell (PIC) simulations of astrophysical plasmas, implemented in a code framework ready for production runs. While explicit PIC methods have gained widespread recognition in the astrophysical community as a reliable tool to simulate plasma phenomena, implicit methods have been seldom explored. This is partly due to the lack of a reliable relativistic implicit PIC formulation that is applicable to state-of-the-art simulations. We propose the RelSIM to fill this gap: our new method is relatively simple, being free of nonlinear iterations and only requiring a global linear solve of the field equations. {With a set of one- and two-dimensional tests}, we demonstrate that the RelSIM produces more accurate results with much smaller numerical errors in the total energy than standard explicit PIC, particularly when characteristic plasma scales (skin depth and plasma frequency) are heavily underresolved on the numerical grid. By construction, the RelSIM also performs much better than the Relativistic Implicit-Moment Method (RelIMM), originally proposed for semi-implicit PIC simulations in the relativistic regime. Our results are promising to conduct large-scale (in terms of duration and domain size) PIC simulations of astrophysical plasmas, potentially reaching physical regimes inaccessible by standard explicit PIC codes.
\end{abstract}



\section{Introduction and review of the particle-in-cell panorama}
\label{sec:intro}
Modern astrophysical research is tightly linked with numerical simulations carried out on supercomputers. In many cases, pen-and-paper calculations do not suffice to analyze the behavior of complex astrophysical systems, whose dynamics is often highly nonlinear and multiphysics, multiscale in nature. Of particular importance in astrophysics is the modeling of plasmas, which are ubiquitous in the Universe. In several astrophysical environments (e.g.\ the surroundings of black holes and neutron stars, in supernova remnants, X-ray binary systems, etc.), plasma dynamics interacts with strong gravity, radiation physics, QED effects, and electromagnetic fields of extreme strengths. This interaction often results in strong plasma-energization processes, through which plasma particles (e.g.\ electrons, positrons, and ions) can be accelerated to relativistic energies. Nonlinear plasma dynamics, particularly coupled with such effects, is hard to describe analytically, resulting in an ever-growing need for advanced simulation tools.

The Particle-in-Cell (PIC) method is among the most successful approaches for the numerical simulation of relativistic astrophysical plasmas. With their origin dating back to the 1960s, PIC methods have attracted widespread attention after the birth of high-performance computing. Nowadays, massively parallel PIC codes are routinely employed for the study of plasma phenomena, particularly in the collisionless regime where binary particle-particle encounters are {negligibly rare}, and particles only interact via self-generated electromagnetic fields. Other computational approaches to model collisionless plasmas exist, but they typically rely on specific assumptions that discard small-scale physics (e.g.\ hybrid methods) or they involve prohibitive computational costs in most practical cases (e.g.\ Vlasov methods).

Even after decades of evolution, the basic structure of a PIC code remains as described in classical textbooks (e.g.\ \citealt{birdsalllangdon1991}): a computational grid is employed to solve a set of Maxwell's equations {(employing CGS units here and in the remainder of the text)},
\be
\frac{\pd\vecE}{\pd t} = c\grad\btimes\vecB - 4\pi\vecJ,
\label{eq:ampere}
\ee
\be
\frac{\pd\vecB}{\pd t} = -c\grad\btimes\vecE,
\label{eq:faraday}
\ee
\be
\grad\bcdot\vecE = 4\pi\rho,
\label{eq:gauss_E}
\ee
\be
\grad\bcdot\vecB = 0,
\label{eq:gauss_B}
\ee
where $c$ is the speed of light, $\vecE$ and $\vecB$ are the electric and magnetic fields, and the current and charge density $\vecJ$ and $\rho$ are source terms linked to the particle motion. In a PIC code, a large number of computational particles are evolved according to the relativistic equations of motion,
\be
\frac{\rmd\vecx}{\rmd t} = \vecv,
\ee
\be
\frac{\rmd\vecu}{\rmd t} = \frac{q}{m} \left(\vecE + \frac{\vecv}{c}\btimes\vecB\right),
\ee
where $\vecx$ and $\vecu$ are the particle position and {the spatial part of the 4-velocity (i.e.\ a 3-vector)}, $q$ and $m$ are the particle charge and mass, and $\vecv=\vecu/\gamma$ with $\gamma=\sqrt{1+u^2/c^2}=1/\sqrt{1-v^2/c^2}$ the relativistic Lorentz factor. Solving the equations of motion with a finite number of particles is essentially equivalent to sampling the particle distribution function $f(\vecx,\vecu,t)$ (i.e.\ a solution of the Vlasov equation) with a Monte Carlo approach; in this sense, $\vecJ$ and $\rho$ entering Maxwell's equations are moments of $f$ gathered from particle quantities onto the grid. Because the number of particles in a PIC simulations is limited by computational resources, the distribution function in PIC runs is (sometimes heavily) affected by numerical noise. Nevertheless, PIC codes have become the primary tool for investigating astrophysical plasmas from first principles, owing to their simplicity, reliability and remarkable performance on parallel computing architectures.

PIC codes employed in the astrophysical community can be divided in two main categories, i) ``explicit'' codes and ii) ``implicit'' codes. Algorithmically, explicit methods involve an explicit discretization of the field equations, which are therefore solved explicitly, followed by a particle ``push'' based on an implicit discretization of the equations of motion which can however be recast into an explicit solution procedure. The explicit approach for fields and particles can vary in flavor, but it typically consists of a temporal leap-frogging procedure where fields and particles are decoupled. This approach is easy to implement and extremely versatile in terms of parallelization and performance. Several codes employed in astrophysical and laboratory plasma research (e.g.\ \textsc{EPOCH}, \citealt{arber2015}; \textsc{OSIRIS}, \citealt{fonseca2002}; \textsc{SHARP}, \citealt{shalaby2017}; \textsc{Smilei}, \citealt{derouillat2018}; \textsc{Tristan v2}, \citealt{hakobyan2023zeltron}; \textsc{VORPAL}, \citealt{nietercary2004}; VPIC, \citealt{bird2022}; \textsc{WarpX}, \citealt{fedeli2022}; \textsc{Zeltron}, \citealt{cerutti2013,bacchini2022mri}; etc.\footnote{The author apologizes for any omissions from this list, which could grow infinitely long in principle.}) have been employing the explicit-PIC approach for a long time, and have successfully attacked many open problems in relativistic astrophysics (e.g.\ \citealt{spitkovsky2008,zhdankin2017,comissosironi2018,guo2021,werneruzdensky2021,sironi2022}), even including quantum-electrodynamics and strong-gravity effects (e.g.\ \citealt{parfrey2019,crinquand2020,sridhar2021,elmellah2022,galishnikova2023,groselj2023,hakobyan2023}), or frame transformations appropriate e.g.\ for expanding/shearing plasmas (\citealt{riquelme2012,hoshino2015,sironinarayan2015,bacchini2022mri,tran2023}). The simplicity of explicit PIC also allows for efficient implementation on new architectures such as GPUs (e.g.\ \textsc{PIConGPU}, \citealt{burau2010}; \textsc{Entity}, Hakobyan et al. 2023, in prep).

Despite their widespread success, explicit codes also suffer from severe limitations linked to numerical instabilities: the explicit discretization introduces artificial unstable modes that essentially destroy the simulation results very rapidly when certain criteria are not met. In particular, locally underresolving temporal and spatial scales such as (the inverse of) the plasma frequency $\omp=\sqrt{4\pi q^2 n/m}$ (where $n$ is the plasma number density) and skin depth $c/\omp$ in many cases produces unphysical results. If these stability conditions are violated even in one single computational cell, the entire simulation may be irremediably compromised. Depending on the physical case, other scales may require full resolution on the grid, e.g.\ the Debye length $\lambda_\mathrm{D}=\sqrt{kT/(4\pi q^2 n)}$ (where $T$ is the plasma temperature). In many cases, these restrictions are not problematic, because the phenomena of interest take place precisely at the aforementioned spatiotemporal scales which therefore need to be resolved to accurately capture the corresponding processes. However, when this is not the case, severe limitations arise on the applicability of explicit codes, where the time step and the grid spacing are determined by the most restrictive plasma conditions in the whole simulation domain. For example, several problems involve a very large separation of scales in which only the phenomena at the largest scales are important, and the smaller scales could in principle be left underresolved; explicit methods require instead to resolve all scales. Similarly, when large density gradients are involved, the local skin depth and inverse plasma frequency could vary dramatically within the domain; interesting physics may take place only in very localized regions where $c/\omp$ and $\omp^{-1}$ are very small, hence only those regions would require finer grids and smaller time steps, but explicit methods will instead impose restrictive simulation parameters everywhere (e.g.\ in compact-object magnetospheres, \citealt{ceruttibeloborodov2016,hakobyan2023}). Finally, when considering the presence of multiple plasma species, restrictions in the numerical parameters arise due to the need to resolve the scales of the lighter species (usually electrons) whereas many interesting phenomena primarily involve the large scales determined by the heavier species (usually ions), resulting in extremely intensive computations (e.g.\ ion-scale magnetic reconnection, large-scale wave decay, solar-wind turbulence, and shocks; \citealt{spitkovsky2008,werner2018,verscharen2019,bacchini2022kaw}). Especially in the latter case, it may be undesirable to simply switch to a different paradigm (e.g.\ hybrid codes, \citealt{caprioli2018,bott2021,squire2022}), since doing so implies discarding potentially interesting electron physics that can still occur at ion scales. In short, explicit PIC codes may simply not suffice, in specific cases of interest, to carry out simulations over large spatial and temporal scales due to an intrinsic limitation of the numerical method that results in prohibitive computing costs.

For these reasons, extensive research has been dedicated to developing implicit PIC methods. These methods do not suffer from the instabilities affecting explicit PIC, and can in principle allow for simulations where spatiotemporal scales are arbitrarily underresolved\footnote{Note that the physics occurring at underresolved scales is not captured accurately, but is rather averaged over.}. Implicit PIC codes essentially involve an implicit discretization of Maxwell's equations, together with a particle push which may or may not be nonlinearly coupled to the field-solver step. If this nonlinear coupling is retained, the resulting approach is usually labeled ``fully implicit'' (\citealt{LapentaMarkidis2011,markidisLapenta2011,bacchini2019fipic,chen2020,angus2023}) and may involve the solution of a very large, nonlinear system of equations, whose dimension can be of the order of the total number of particles in a simulation. Such extremely large systems are hard to handle in practice, since convergence of iterative solution methods is not guaranteed; even with advanced preconditioning, it is not straightforward to obtain acceptable scaling behavior on supercomputing infrastructures. Several approaches have been developed to ameliorate the problem, e.g.\ the reduction of the nonlinear system via nonlinear substitution of the particle equations into the field equations {(``kinetic enslavement'', e.g.\ \citealt{markidisLapenta2011,taitano2013,bacchini2019fipic} and references therein)}. Even with such improvements, fully implicit PIC codes have not reached a level of maturity that makes them applicable in practical situations.

When the particle push and the field advance are decoupled in implicit PIC methods, these are termed ``semi-implicit'' and present a much lower level of complexity with respect to fully implicit PIC methods. The decoupling essentially consists of rewriting the source terms in Maxwell's equations as linear functions of the electromagnetic fields, thereby removing any nonlinearity and reducing the problem to a linear-solve step on the grid. This decoupling can be carried out in several fashions (see next Sections), either approximately (i.e.\ using a linearization) via the so-called Implicit-Moment Method (IMM, e.g.\ \citealt{brackbillforslund1982,lapenta2006}) or exactly in the case of the Energy-Conserving Semi-Implicit Method (ECSIM, \citealt{lapenta2017ecsimcode}). The latter is particularly interesting because, as the name suggests, solving the implicit equations without approximations results in the \emph{exact} (i.e.\ to machine precision) conservation of total energy throughout the numerical simulation, a feat that no currently employed explicit method achieves. Although the original method has been later refined and improved (e.g.\ \citealt{chentoth2019,campospintopages2022}), conservation of energy is particularly important for stability, and the ECSIM has demonstrated the capability to allow for very long simulations on very large spatial scales (e.g.\ \citealt{park2019,zhou2019,arro2022}; Pezzini et al.\ 2023, in prep.). The important caveat here is that the IMM and ECSIM are \emph{nonrelativistic}, i.e.\ by assuming that particle speeds are much smaller than $c$, the particle-push step is replaced with its nonrelativistic counterpart where Lorentz factors are unitary and the equations of motion simplify to the Newtonian limit. In the relativistic regime instead, constructing a semi-implicit PIC method is more involved precisely due to the presence of the Lorentz factor, which introduces an intrinsic nonlinearity in the particle equations of motion. This detail is crucial: a relativistic version of the IMM (termed RelIMM, \citealt{Noguchi2007,kempf2015}) can be formulated, but its applicability is severely hindered by the nonlinearity of the particle equations (see Section~\ref{sec:RelIMM}). The ECSIM, instead, simply cannot be directly extended to relativistic applications while also retaining its exact energy-conservation properties, because the reformulation of the Maxwell sources into linear functions of the fields cannot be carried out without approximations in the relativistic case (see Section~\ref{sec:RelSIM}). As a consequence, the only relativistic semi-implicit PIC method presented in literature so far is the aforementioned RelIMM, which however performs poorly in practical applications (see Section~\ref{sec:tests}) and has never been applied in production runs.

The focus of this work is a novel, simple, and reliable semi-implicit PIC method that is ready for production simulations of astrophysical plasma phenomena in the relativistic regime. We call the new method the Relativistic Semi-Implicit Method (RelSIM); like the ECSIM, the RelSIM retains a simple formulation based on ``mass matrices'' (see  Section~\ref{sec:RelSIM}), is free of nonlinear iterations, and surpasses the RelIMM in terms of performance and quality of the results (see Section~\ref{sec:tests}). Due to the intrinsic nonlinear nature of the relativistic Vlasov-Maxwell system, our approach to remove nonlinear iterations necessarily sacrifices exact energy conservation; however, we demonstrate that the new RelSIM still possesses excellent energy-conservation properties, which make it superior to the RelIMM. Because the new method is implicit in nature, it can be employed in situations where plasma scales are dramatically underresolved without loss of stability, in contrast with explicit methods.

This work is organized as follows: in Section~\ref{sec:RelIMM} we review (and improve upon, with a generalized reformulation) the RelIMM originally presented in \cite{Noguchi2007}. In Section~\ref{sec:RelSIM} we derive and present the new RelSIM. In Section~\ref{sec:tests} we present quantitative comparisons between relativistic explicit and semi-implicit PIC methods (including the new RelSIM) in a number of representative test cases. Finally, in Section~\ref{sec:conclusions} we discuss and summarize our results.

\section{Review of the relativistic implicit-moment method}
\label{sec:RelIMM}
In this Section we review the RelIMM (\citealt{Noguchi2007,kempf2015}) to which we add modifications and improvements. The original RelIMM is based on a $\theta$-scheme applied to Maxwell's equations for each grid element $g$,
\be
\frac{\vecE_g^{n+\theta}-\vecE_g^n}{\theta\Delta t} = c\grad\btimes\vecB_g^{n+\theta} - 4\pi\vecJ_g^{n+1/2},
\label{eq:ampere_disc}
\ee
\be
\frac{\vecB_g^{n+\theta}-\vecB_g^n}{\theta\Delta t} = -c\grad\btimes\vecE_g^{n+\theta},
\label{eq:faraday_disc}
\ee
\be
\grad\bcdot\vecE_g^{n+\theta} = 4\pi\rho_g^{n+\theta},
\label{eq:gauss_disc}
\ee
where electromagnetic fields at $n+\theta$ are calculated as a linear interpolation between integer temporal steps, e.g.\ $\vecE^{n+\theta} = \theta\vecE^{n+1}+(1-\theta)\vecE^n$ with $\theta\in[1/2,1]$. The condition $\grad\bcdot\vecB=0$ is automatically satisfied at all times if the computational grid possesses mimetic properties {(i.e.\ it preserves the basic analytic properties of differential operators, see e.g.\ \citealt{lipnikov2014} for a review)}. For each particle $p$, the relativistic equations of motion are
\be
\frac{\vecx_p^{n+1}-\vecx_p^n}{\Delta t} = \vecvbar_p,
\label{eq:pos_discrete}
\ee
\be
\frac{\vecu_p^{n+1}-\vecu_p^n}{\Delta t} = \frac{q_p}{m_p}\left(\vecE^{n+\theta}(\vecx_p^{n+1/2}) + \frac{\vecvbar_p}{c}\btimes\vecB^n(\vecx_p^{n+1/2})\right),
\label{eq:mom_discrete}
\ee
where the half-step particle position $\vecx^{n+1/2} = (\vecx^{n+1}+\vecx^n)/2$, and $\vecvbar$ is an arbitrarily defined half-step velocity. The precise definition of $\vecvbar$ is what distinguishes different particle pushers. In the nonrelativistic regime (i.e.\ $\vecu=\vecv$), the unambiguous definition $\vecvbar\equiv(\vecv^{n+1}+\vecv^n)/2$ provides second-order accuracy and allows for solving the momentum equation~\eqref{eq:mom_discrete} with a simple operator-split approach (the ``Boris'' method, \citealt{boris1970}). In the relativistic regime, defining $\vecvbar$ is nonstraighforward due to the nonlinearity in the Lorentz factor  (see \citealt{ripperda2018a} for a review on the subject). Several definitions of $\vecvbar$ have been presented in literature (e.g.\ \citealt{boris1970,vay2008,LapentaMarkidis2011,higueracary2017}); in most cases, the half-step velocity is of the form
\be
\vecvbar = \frac{\vecu^{n+1}+\vecu^n}{2\bgam},
\label{eq:vbarrel}
\ee
such that no work is exerted on computational particles by magnetic fields\footnote{This can be seen by dotting eqs.~\eqref{eq:vbarrel} and \eqref{eq:mom_discrete} and noticing that only electric fields contribute to a particle's change in energy.}, reflecting reality. This is verified regardless of the definition of $\bgam$, which then acts as the true discriminant between relativistic particle pushers. A popular choice is the relativistic Boris pusher, where
\be
\bgam = \sqrt{1+\left[\vecu^n+q\Delta t\vecE^{n+\theta}(\vecx_p^{n+1/2})/(2m)\right]^2/c^2},
\label{eq:bgam_boris}
\ee
which allows the direct solution of eq.~\eqref{eq:mom_discrete} since $\bgam$ can be computed from known quantities (if particle positions and electromagnetic fields are known). Another option, considered in the original RelIMM algorithm (\citealt{Noguchi2007}), and later discussed in detail by \cite{LapentaMarkidis2011}, is the definition
\be
\bgam = (\gamma^{n+1}+\gamma^n)/2,
\label{eq:bgam_LM}
\ee
which possesses important properties for energy conservation: with this $\bgam$, it is straightforward to show that
\be
\ba
m_p\vecvbar_p\bcdot(\vecu_p^{n+1}-\vecu_p^n) & = m_p c^2(\gamma_p^{n+1}-\gamma_p^n) \\
& = \vecvbar_p\Delta t\bcdot q_p\vecE^{n+\theta}(\vecx_p^{n+1/2}) \\
& = (\vecx_p^{n+1}-\vecx_p^n)\bcdot q_p\vecE^{n+\theta}(\vecx_p^{n+1/2}),
\ea
\ee
resulting in the physically correct consequence that the change in a particle's kinetic energy $mc^2\gamma$ between two time steps is exactly equal to the work done by the electric field during that time step. Other choices of $\bgam$ do not respect this condition. However, choosing the half-step Lorentz factor~\eqref{eq:bgam_LM} implies a much more complicated solution of the momentum equation~\eqref{eq:mom_discrete} than in the Boris case (see Section~\ref{sec:RelIMM_particles} and Appendix~\ref{app:LMsolution}). Ultimately, the RelIMM can be formulated with any sensible choice of $\bgam$, as we will show later.

\subsection{RelIMM: Field Solver}
\label{sec:RelIMM_fields}
To construct a RelIMM scheme from our discretized Maxwell's equations, we start by combining Amp\`{e}re's and Faraday's laws: by taking the curl of eq.~\eqref{eq:faraday_disc} and inserting it into eq.~\eqref{eq:ampere_disc} we get
\be
\ba
\vecE_g^{n+\theta} & + (c\theta\Delta t)^2\grad\btimes\grad\btimes\vecE_g^{n+\theta} \\
& = \vecE_g^n + c\theta\Delta t\grad\btimes\vecB_g^n - 4\pi \theta\Delta t\vecJ_g^{n+1/2}.
\ea
\label{eq:ampere_disc_2ndorder}
\ee
Then, we expand the curl term $\grad\btimes\grad\btimes\vecE^{n+\theta}=\grad\grad\bcdot\vecE^{n+\theta}-\grad^2\vecE^{n+\theta}$ and use Gauss's law~\eqref{eq:gauss_disc} to obtain
\be
\ba
\vecE_g^{n+\theta} & - (c\theta\Delta t)^2\grad^2\vecE_g^{n+\theta} \\
& = \vecE_g^n + c\theta\Delta t\grad\btimes\vecB_g^n - 4\pi \theta\Delta t\vecJ_g^{n+1/2} \\ 
& \quad - 4\pi(c\theta\Delta t)^2\grad\rho_g^{n+\theta}.
\label{eq:ampere_disc_2ndorder_expanded}
\ea
\ee
Finally, we can employ a ($\theta$-scheme-discretized) charge-continuity equation $\pd\rho/\pd t = -\grad\bcdot\vecJ$ to express the charge density at $n+\theta$ as a function of the half-step current,
\be
\rho_g^{n+\theta} = \rho_g^n - \theta\Delta t\grad\bcdot\vecJ_g^{n+1/2},
\ee
which inserted into the previous equation gives
\be
\ba
\vecE_g^{n+\theta} & - (c\theta\Delta t)^2\grad^2\vecE_g^{n+\theta} \\
& = \vecE_g^n + c\theta\Delta t\grad\btimes\vecB_g^n - 4\pi \theta\Delta t\vecJ_g^{n+1/2} \\ 
& \quad - 4\pi(c\theta\Delta t)^2\grad\rho_g^n + 4\pi c^2(\theta\Delta t)^3 \grad\grad\bcdot\vecJ_g^{n+1/2}.
\label{eq:IMM_field_1}
\ea
\ee

Eq.~\eqref{eq:IMM_field_1} is the central point of interest of the IMM. In principle, the sources and specifically the half-step current $\vecJ^{n+1/2}$ provide a nonlinear coupling of Maxwell's equations with the particle equations of motion: the current is defined as
\be
\vecJ_g^{n+1/2} = \frac{1}{\Delta V_g}\sum_{p} q_p\vecvbar_p W(\vecx_p^{n+1/2}-\vecx_g),
\label{eq:IMM_jbar}
\ee
where $\Delta V_g$ is the volume associated with each grid element and $W(\vecx_p^{n+1/2}-\vecx_g)$ is a chosen interpolation function (usually a first-order b-spline). Note that here $\vecvbar$ is a function of $\vecE^{n+\theta}$ via eq.~\eqref{eq:mom_discrete}, and $\vecx^{n+1/2}$ is a function of $\vecvbar$ (and thus of $\vecE^{n+\theta}$) via eq.~\eqref{eq:pos_discrete}. Because of this, eq.~\eqref{eq:IMM_field_1} and the particle equations of motion in principle constitute a very large (of size $\sim N_p$), fully coupled nonlinear system to be solved in order to advance the numerical solution. The core of the IMM approach consists instead of solving a \emph{linear} system to find $\vecE^{n+\theta}$, by recasting $\vecJ^{n+1/2}$ as a linear function of the electric field. To do so, we first expand $W$ around $\vecx_p^{n+1/2}$,
\be
\ba
& W(\vecx_p^{n+1/2}-\vecx) = \\
& W(\vecx_p^n-\vecx) - (\vecx_p^{n+1/2}-\vecx_p^n)\bcdot\grad W(\vecx_p^n-\vecx) \\
& +\frac{1}{2}(\vecx_p^{n+1/2}-\vecx_p^n)(\vecx_p^{n+1/2}-\vecx_p^n)\bdbldot\grad\grad W(\vecx_p^n-\vecx) \\
& + \dots
\ea
\label{eq:Wexpansion}
\ee
and recognizing $\vecx_p^{n+1/2}-\vecx_p^n=(\Delta t/2)\vecvbar_p$, we can substitute in the expression for the current~\eqref{eq:IMM_jbar} keeping terms up to first order in $\Delta t$,
\be
\ba
\vecJ_g^{n+1/2} & = \frac{1}{\Delta V_g}\sum_p q_p\vecvbar_p W(\vecx_p^{n+1/2}-\vecx_g) \\ 
& - \frac{\Delta t}{2\Delta V_g}\grad\bcdot\sum_p q_p \vecvbar_p\vecvbar_p W(\vecx_p^{n+1/2}-\vecx_g) \\
& + \mathcal{O}(\Delta t^2),
\label{eq:IMM_jbar_1}
\ea
\ee
where we have also used vector identities to bring the divergence operator out of the summation. In this way, we have removed the nonlinear dependence of the interpolation function on $\vecvbar$ (and therefore on $\vecE^{n+\theta}$). Next, we will construct a \emph{linear} dependence of the velocity on the unknown electric field. We consider the momentum equation~\eqref{eq:mom_discrete} and assume a definition of the half-step velocity $\vecvbar=(\vecu^{n+1}+\vecu^n)/(2\bgam)$,
\be
\bgam_p\vecvbar_p = \vecu_p^n + \frac{q_p\Delta t}{2m_p}\left(\vecE^{n+\theta}(\vecx_p^{n+1/2}) + \frac{\vecvbar_p}{c}\btimes\vecB^n(\vecx_p^{n+1/2})\right).
\ee
This equation could be easily solved for $\vecvbar$ if $\bgam$ were known. However, the latter is in the most general case a nonlinear function of $\vecvbar$, which prevents the formal inversion of the equation above. In addition, to construct $\vecJ^{n+1/2}$ as a linear function of $\vecE^{n+\theta}$, we need $\vecvbar$ itself to be a linear function of $\vecE^{n+\theta}$. For this reason we are forced to introduce an approximation where we replace the unknown $\bgam\simeq\Gamma$, such that 
\be
\Gamma_p\vecvbar_p = \vecu_p^n + \frac{q_p\Delta t}{2m_p}\left(\vecE^{n+\theta}(\vecx_p^{n+1/2}) + \frac{\vecvbar_p}{c}\btimes\vecB^n(\vecx_p^{n+1/2})\right),
\ee
and we assume that $\Gamma$ can be computed explicitly without knowing $\bgam$. The definition of $\Gamma$ depends on the chosen particle pusher: for the Boris pusher, we can approximate eq.~\eqref{eq:bgam_boris} as
\be
\bgam_p\simeq\Gamma_p \equiv \sqrt{1+\left[\vecu_p^n+q_p\Delta t\vecE^{n}(\vecx_p^n)/(2m_p)\right]^2/c^2},
\ee
using only known quantities to calculate $\Gamma$. Likewise, for the Lapenta-Markidis definition~\eqref{eq:bgam_LM},
\be
\bgam_p\simeq\Gamma_p \equiv \gamma_p^n + \frac{q_p\Delta t}{2m_pc^2}\vecE^n(\vecx_p^n)\bcdot\vecv_p^n,
\label{eq:approxGammaLM}
\ee
and so on for other choices of particle pushers. Our formulation here provides a more general version of the original RelIMM (\citealt{Noguchi2007}), since here we allow for choices of pushers other than the Lapenta-Markidis one. With the approximation introduced via $\Gamma$, we can write an explicit solution of eq.~\eqref{eq:mom_discrete},
\be
\vecvbar_p = \bb{\alpha}_p\vecu_p^n + \frac{q_p\Delta t}{2m_p} \bb{\alpha}_p\vecE_p^{n+\theta},
\label{eq:vbar_imm}
\ee
where
\be
\bb{\alpha}_p = \frac{1}{\Gamma_p(1+\beta_p^2)}\left[\mathbb{I}-\mathbb{I}\btimes\vecbeta_p/\Gamma_p+\vecbeta_p\vecbeta_p/\Gamma_p^2\right],
\ee
with $\vecbeta_p=q_p\Delta t\vecB_p^n/(2m_p c)$ and we have used the shorthand notation $\vecE_p^{n+\theta} = \vecE^{n+\theta}(\vecx_p^{n+1/2})$, $\vecB_p^{n} = \vecB^{n}(\vecx_p^{n+1/2})$. Inserting eq.~\eqref{eq:vbar_imm} into eq.~\eqref{eq:IMM_jbar_1} and keeping first-order terms yields
\be
\ba
\vecJ_g^{n+1/2} & \simeq \frac{1}{\Delta V_g}\sum_p q_p\bb{\alpha}_p\vecu_p^n W(\vecx_p^n-\vecx_g) \\ 
& + \frac{\Delta t}{2\Delta V_g}\sum_p \frac{q_p^2}{m_p}\bb{\alpha}_p\vecE_p^{n+\theta} W(\vecx_p^n-\vecx_g) \\ 
& - \frac{\Delta t}{2\Delta V_g}\grad\bcdot\sum_p q_p (\bb{\alpha}_p\vecu_p^n)(\bb{\alpha}_p\vecu_p^n) W(\vecx_p^n-\vecx_g) \\
& + \mathcal{O}(\Delta t^2).
\label{eq:IMM_jbar_2}
\ea
\ee
To finally obtain an expression for $\vecJ^{n+1/2}$ as a linear function of $\vecE^{n+\theta}$, we need to introduce further approximations. First,  since $\vecx_p^{n+1/2}$ needed to evaluate $\vecE_p^{n+\theta}$ and $\vecB_p^{n}$ is not known when calculating the current, we employ
\be
\vecB_p^{n} \simeq \vecB^{n}(\vecx_p^{n}) = \sum_{g'}\vecB_{g'}^n W(\vecx_p^{n}-\vecx_{g'}),
\ee
\be
\vecE_p^{n+\theta} \simeq \vecE^{n+\theta}(\vecx_p^{n}) = \sum_{g'}\vecE_{g'}^{n+\theta} W(\vecx_p^{n}-\vecx_{g'}).
\ee
Second, we bring the electric field out of the summation,
\be
\ba
\sum_p & \frac{q_p^2}{m_p} \bb{\alpha}_p \left(\sum_{g'}\vecE_{g'}^{n+\theta} W(\vecx_p^{n}-\vecx_{g'})\right) W(\vecx_p^n-\vecx_g) \\
& \simeq \left(\sum_p \frac{q_p^2}{m_p}\bb{\alpha}_p W(\vecx_p^{n}-\vecx_g)\right) \vecE_g^{n+\theta}.
\ea
\label{eq:approxNGPE}
\ee
This equality is exactly true when $W$ is a zeroth-order b-spline (i.e.\ the interpolation is of nearest-grid-point type). This choice of $W$ is very uncommon as it introduces high levels of noise in the interpolated data. For the more common choice of first-order b-splines, the operation above introduces a (rather crude) approximation, which as we will show results in artificial energy damping. This choice however is functional to obtain a final expression of the current that solely requires particle quantities at the previous time step and that is linear in the unknown electric field,
\be
\ba
\vecJ_g^{n+1/2} & \simeq \frac{1}{\Delta V_g}\sum_p q_p\bb{\alpha}_p\vecu_p^n W(\vecx_p^n-\vecx_g) \\ 
& + \frac{\Delta t}{2\Delta V_g}\left(\sum_p \frac{q_p^2}{m_p}\bb{\alpha}_p W(\vecx_p^{n}-\vecx_g)\right)\vecE_g^{n+\theta} \\ 
& - \frac{\Delta t}{2\Delta V_g}\grad\bcdot\sum_p q_p (\bb{\alpha}_p\vecu_p^n)(\bb{\alpha}_p\vecu_p^n) W(\vecx_p^n-\vecx_g).
\label{eq:IMM_jbar_3}
\ea
\ee
Inserting eq.~\eqref{eq:IMM_jbar_3} into eq.~\eqref{eq:IMM_field_1} yields the final field equation of the RelIMM,
\be
\ba
(\mathbb{I} & +\bb{\mu}_g)\vecE_g^{n+\theta} - (c\theta\Delta t)^2 \left[\grad^2\vecE_g^{n+\theta} + \grad\grad\bcdot(\bb{\mu}_g\vecE_g^{n+\theta})\right] \\
& = \vecE_g^n + \theta\Delta t \left[c\grad\btimes\vecB_g^n - 4\pi\left(\widehat{\vecJ}_g - \frac{\Delta t}{2}\grad\bcdot\widehat{\bb{\Pi}}_g\right)\right] \\ 
& \quad - 4\pi(c\theta\Delta t)^2 \left[\grad\rho_g^n - \theta\Delta t \grad\bcdot\left(\widehat{\vecJ}_g - \frac{\Delta t}{2}\grad\bcdot\widehat{\bb{\Pi}}_g\right)\right],
\label{eq:IMM_field_2}
\ea
\ee
where 
\be
\bb{\mu}_g = \frac{2\pi\theta\Delta t^2}{\Delta V_g}\sum_p \frac{q_p^2}{m_p}\bb{\alpha}_pW(\vecx_p^n-\vecx_g),
\ee
\be
\widehat{\vecJ}_g = \frac{1}{\Delta V_g}\sum_p q_p\bb{\alpha}_p\vecu_p^n W(\vecx_p^n-\vecx_g),
\ee
\be
\widehat{\bb{\Pi}}_g = \frac{1}{\Delta V_g}\sum_p q_p (\bb{\alpha}_p\vecu_p^n)(\bb{\alpha}_p\vecu_p^n) W(\vecx_p^n-\vecx_g).
\ee
Eq.~\eqref{eq:IMM_field_2} is linear in $\vecE^{n+\theta}$ and thus can be solved efficiently with any standard linear solver, once the source terms have been calculated from known particle quantities. The magnetic field can then be updated using eq.~\eqref{eq:faraday_disc} and extrapolating to $\vecB^{n+1}=(\vecB^{n+\theta}-(1-\theta)\vecB^n)/\theta$.

\subsection{RelIMM: Particle Push}
\label{sec:RelIMM_particles}
Once eq.~\eqref{eq:IMM_field_2} is solved on the grid, the particles can be evolved according to eqs.~\eqref{eq:pos_discrete} and \eqref{eq:mom_discrete}. Because the two equations are nonlinearly coupled ($\vecx^{n+1/2}$ depends on $\vecvbar$ and vice versa), the particle push is carried out iteratively, starting from an initial guess for $\vecvbar$, according to the following steps:
\begin{enumerate}
    \item Compute $\vecx^{n+1}$ (and thus $\vecx^{n+1/2}$) using the current $\vecvbar$;
    \item Interpolate $\vecE^{n+\theta}$ and $\vecB^n$ from the grid onto the current particle position $\vecx^{n+1/2}$;
    \item Compute $\bgam$ according to the preferred definition (e.g.\ Boris, Lapenta-Markidis, etc.);
    \item Compute $\vecu^{n+1}$ (and thus $\vecvbar$) using $\bgam$ and the interpolated fields.
\end{enumerate}
Step 3 above is carried out differently for different particle pushers. For the Lapenta-Markidis pusher (\citealt{LapentaMarkidis2011}), however, no explicit expression for $\bgam$ has been presented in literature, to the best of our knowledge. In Appendix~\ref{app:LMsolution} we report such an explicit solution for the first time. Once $\bgam$ is known, we can obtain the new particle 4-velocity as $\vecu^{n+1}=2\bgam\vecvbar-\vecu^n$, assuming the half-step velocity has been defined as $\vecvbar=(\vecu^{n+1}+\vecu^n)/(2\bgam)$. The iterative solution of the particle equations of motion can then continue as illustrated above. In typical implementations of the (Rel)IMM, this iteration is not carried out until convergence {(which is not guaranteed due to the nonlinearly implicit nature of the equations)}, but rather steps 1--4 above are repeated a fixed number of times to avoid excessive computational costs.

\subsection{Summary of the RelIMM}
One complete time iteration of the RelIMM is composed of the following steps:
\begin{enumerate}
    \item Gather the source terms $\rho^n$, $\widehat{\vecJ}$, $\widehat{\bb{\Pi}}$ on the grid from known particle quantities at time step $n$.
    \item Solve eq.~\eqref{eq:IMM_field_2} for $\vecE^{n+\theta}$ using any preferred linear solver.
    \item Update the position and 4-velocity of all particles iteratively by solving the coupled system \eqref{eq:pos_discrete}--\eqref{eq:mom_discrete} with the preferred definition of $\bgam$ and using $\vecE^{n+\theta}$ and $\vecB^n$.
    \item Finalize the field solution on the grid by computing $\vecE^{n+1}$ and $\vecB^{n+1}$.
\end{enumerate}

Although this formulation of the RelIMM is rather simple as it only involves a linear solve for the grid quantities, it also presents several drawbacks and approximations:
\begin{itemize}
    \item {To interpolate source quantities from particles to grid, the interpolation function is expanded around $\vecx_p^{n+1/2}$, since $\vecx_p^{n+1/2}$ is not known when gathering the sources (see eq.~\eqref{eq:Wexpansion})}. This results in the need to calculate an additional source quantity $\widehat{\bb{\Pi}}$.
    \item In the definition of the rotation matrix $\bb{\alpha}$, $\bgam$ is approximated to an expression (which depends on the chosen pusher) such that it can be evaluated using known field and particle quantities at the previous time step (see e.g.\ eq.~\eqref{eq:approxGammaLM}).
    \item Crucially, to make the dependence of $\vecJ^{n+1/2}$ on $\vecE^{n+\theta}$ linear, it is assumed that the particle-grid interpolation functions are zeroth-order b-splines (see eq.~\eqref{eq:approxNGPE}), which is rarely the case due to excessive numerical noise introduced by low-order interpolation.
    \item Since the particle position and 4-velocity are synchronized in time, the particle-push step involves an iteration that needs to be carried out until convergence (in principle, although in practice a fixed number of iterations are realized instead).
\end{itemize}
The performance of the standard RelIMM, even compared to an explicit relativistic PIC method, is severely affected by the shortcomings listed above. As we will show, the approximations introduced result in large errors in the total energy when employing coarse grid resolutions. Since reducing the number of cells in relativistic PIC simulations is in principle the main advantage of implicit methods, this particular point renders the standard RelIMM rather unattractive. In the next Section, we present a new method that eliminates many of the drawbacks affecting the RelIMM.

\section{The new RelSIM formulation}
\label{sec:RelSIM}
Here, we present a new Relativistic Semi-Implicit Method (RelSIM) for PIC that substantially improves over the standard RelIMM. Our approach extends the nonrelativistic ECSIM method (\citealt{lapenta2017ecsimcode}) to relativistic regimes, and is free of many of the drawbacks affecting the RelIMM. The new method necessarily sacrifices exact energy conservation in order to discard nonlinear iterations, but energy errors are much smaller than those observed when applying the RelIMM (see Section~\ref{sec:tests}).

We construct the new RelSIM method starting with the same $\theta$-scheme employed in the IMM for the discretized Maxwell's equations,
\be
\frac{\vecE_g^{n+\theta}-\vecE_g^n}{\theta\Delta t} = c\grad\btimes\vecB_g^{n+\theta} - 4\pi\vecJ_g^{n+1/2},
\label{eq:ampere_disc_relsim}
\ee
\be
\frac{\vecB_g^{n+\theta}-\vecB_g^n}{\theta\Delta t} = -c\grad\btimes\vecE_g^{n+\theta},
\label{eq:faraday_disc_relsim}
\ee
and a slightly modified particle pusher,
\be
\frac{\vecx_p^{n+1/2}-\vecx_p^{n-1/2}}{\Delta t} = \frac{\vecu_p^n}{\gamma_p^n},
\label{eq:pos_discrete_relsim}
\ee
\be
\frac{\vecu_p^{n+1}-\vecu_p^n}{\Delta t} = \frac{q_p}{m_p}\left(\vecE^{n+\theta}(\vecx_p^{n+1/2}) + \frac{\vecvbar_p}{c}\btimes\vecB^n(\vecx_p^{n+1/2})\right),
\label{eq:mom_discrete_relsim}
\ee
where the position update is now staggered with respect to the velocity and can be carried out separately from the velocity update. Like for the RelIMM, the choice of $\vecvbar$ is free but we assume it to have the form $\vecvbar=(\vecu^{n+1}+\vecu^n)/(2\bgam)$. The velocity update is then performed by choosing $\bgam$ according to the form given by available particle pushers (Boris, Lapenta-Markidis, etc.).

\subsection{RelSIM: Field Solver}
Next, we derive the field solver of the new RelSIM. We follow the same steps presented in Section~\ref{sec:RelIMM_fields} to recast Maxwell's equations into the form
\be
\ba
\vecE_g^{n+\theta} & + (c\theta\Delta t)^2\grad\btimes\grad\btimes\vecE_g^{n+\theta} \\
& = \vecE_g^n + c\theta\Delta t\grad\btimes\vecB_g^n - 4\pi \theta\Delta t\vecJ_g^{n+1/2},
\ea
\ee
but we now avoid the expansion of the $\grad\btimes\grad\btimes$ term\footnote{{This expansion can be in principle still carried out, but the following substitution $\grad\bcdot\vecE=4\pi\rho$ (which is operated for the RelIMM) would result in a loss of energy conservation.}} that was performed for eq.~\eqref{eq:ampere_disc_2ndorder_expanded}. Next, we employ the expression for the half-step current,
\be
\vecJ_g^{n+1/2} = \frac{1}{\Delta V_g}\sum_{p} q_p\vecvbar_p W(\vecx_p^{n+1/2}-\vecx_g),
\label{eq:RelSIM_jbar}
\ee
which can now be collected from the particles at position $\vecx^{n+1/2}$, computed from the known velocity $\vecu^n$. Here we can directly substitute the (approximate) solution of eq.~\eqref{eq:mom_discrete_relsim} for $\vecvbar$ and separate out to find
\be
\ba
\vecJ_g^{n+1/2} & \simeq \frac{1}{\Delta V_g}\sum_p q_p\bb{\alpha}_p\vecu_p^n W(\vecx_p^{n+1/2}-\vecx_g) \\ 
& + \frac{\Delta t}{2\Delta V_g}\sum_p \frac{q_p^2}{m_p}\bb{\alpha}_p\vecE_p^{n+\theta} W(\vecx_p^{n+1/2}-\vecx_g),
\label{eq:RelSIM_jbar_2}
\ea
\ee
where $\bb{\alpha}$ is again given by 
\be
\bb{\alpha}_p = \frac{1}{\Gamma_p(1+\beta_p^2)}\left[\mathbb{I}-\mathbb{I}\btimes\vecbeta_p/\Gamma_p+\vecbeta_p\vecbeta_p/\Gamma_p^2\right],
\ee
and $\vecbeta_p=q_p\Delta t\vecB_p^n/(2m_p c)$, $\vecE_p^{n+\theta} = \vecE^{n+\theta}(\vecx_p^{n+1/2})$, $\vecB_p^{n} = \vecB^{n}(\vecx_p^{n+1/2})$. Through $\bb{\alpha}$ we have introduced the first (and only) approximation needed to construct our new method, i.e.\ the assumption $\bgam\simeq\Gamma$ with $\Gamma$ defined according to the chosen particle pusher (see Section~\ref{sec:RelIMM}). Now, differently from the IMM approach, we bring the unknown electric field out of the summation over $p$ in a manner that does not introduce any further approximations, i.e.
\be
\ba
\sum_p & \frac{q_p^2}{m_p} \bb{\alpha}_p \left(\sum_{g'}\vecE_{g'}^{n+\theta} W(\vecx_p^{n+1/2}-\vecx_{g'})\right) W(\vecx_p^{n+1/2}-\vecx_g) \\
& = \sum_{g'}\textbf{M}_{gg'}\vecE_{g'}^{n+\theta},
\ea
\label{eq:massmatrixE}
\ee
where
\be
\textbf{M}_{gg'} = \sum_p \frac{q_p^2}{m_p}\bb{\alpha}_p W(\vecx_p^{n+1/2}-\vecx_{g'})W(\vecx_p^{n+1/2}-\vecx_g)
\ee
is the \emph{mass matrix} first introduced by \cite{lapenta2017ecsimcode} for the original nonrelativistic \textsc{ECSIM}. The difference here is the presence of the relativistic Lorentz factor $\Gamma$ in the definition of $\bb{\alpha}$ above. Comparing to the RelIMM (eq.~\eqref{eq:approxNGPE}), we observe that here we are not introducing any assumption on the interpolation functions $W$. For the common choice of first-order b-splines, eq.~\eqref{eq:massmatrixE} requires the calculation of 9 mass matrices (when the electric field has 3 components) per grid point. Inserting the equation above into the definition of the current, we obtain a final expression for the field equation of the RelSIM,
\be
\ba
\vecE_g^{n+\theta} & + (c\theta\Delta t)^2\grad\btimes\grad\btimes\vecE_g^{n+\theta} + \sum_{g'}\bb{\mu}_{gg'}\vecE_{g'}^{n+\theta} \\
& = \vecE_g^n + c\theta\Delta t\grad\btimes\vecB_g^n - 4\pi\theta\Delta t\widehat{\vecJ}_g,
\ea
\label{eq:RelSIM_field_2}
\ee
where
\be
\bb{\mu}_{gg'} = \frac{2\pi\theta\Delta t^2}{\Delta V_g}\textbf{M}_{gg'},
\ee
\be
\widehat{\vecJ}_g = \frac{1}{\Delta V_g}\sum_p q_p\bb{\alpha}_p\vecu_p^n W(\vecx_p^{n+1/2}-\vecx_g).
\ee
Eq.~\eqref{eq:RelSIM_field_2} is linear in $\vecE^{n+\theta}$ and can be handled with a linear solver. Like in the case of the RelIMM, $\vecB^n$ can be updated once the electric field is known.

\subsection{RelSIM: Particle Push}
The particle update is easier for the RelSIM than for the RelIMM, since the former involves no iteration (see Section~\ref{sec:RelIMM_particles}). Because particle positions and velocities are staggered in time, here we need to  update the position via eq.~\eqref{eq:pos_discrete_relsim} before the field solve. Once $\vecE^{n+\theta}$ is known on the grid, the velocity update~\eqref{eq:mom_discrete_relsim} can be carried out according to the preferred relativistic particle pusher.

\subsection{Summary of the RelSIM}
One complete time iteration of the RelSIM is composed of the following steps:
\begin{enumerate}
    \item Update the position of all particles by solving eq.~\eqref{eq:pos_discrete_relsim}.
    \item Compute the current $\widehat{\vecJ}$ and mass matrices $\textbf{M}$ on the grid from known particle quantities ($\vecx^{n+1/2}$ and $\vecu^n$).
    \item Solve eq.~\eqref{eq:RelSIM_field_2} for $\vecE^{n+\theta}$ using any preferred linear solver.
    \item Update the 4-velocity of all particles by solving the momentum equation~\eqref{eq:mom_discrete_relsim} with the preferred definition of $\bgam$ and using $\vecE^{n+\theta}$ and $\vecB^n$.
    \item Finalize the field solution on the grid by computing $\vecE^{n+1}$ and $\vecB^{n+1}$.
\end{enumerate}

In contrast with the RelIMM, the RelSIM does not need any nonlinear iterations for the particle update and does not require the calculation of the dielectric tensor $\widehat{\bb{\Pi}}$; furthermore, the RelSIM does not rely on any assumptions other than an approximation of the half-step Lorentz factor $\Gamma\simeq\bgam$ needed to linearize the field equations. As a drawback, the RelSIM requires the calculation of the mass matrices, which adds to the complexity of the field solver. However, we will show that this downside is largely compensated by the superior energy-conservation properties of the RelSIM with respect to the original RelIMM.

\section{Validation tests}
\label{sec:tests}
In this Section, we perform several test simulations to assess the numerical performance of the new RelSIM. In general, we compare the results obtained with the RelSIM to those obtained with the RelIMM as well as with a standard explicit-PIC code. For the explicit runs, we employ \textsc{Zeltron}, which is a state-of-the-art tool utilized for many production applications in relativistic astrophysics (e.g.\ \citealt{cerutti2013,zhdankin2017,werner2018,parfrey2019,mehlhaff2021,bacchini2022mri,galishnikova2023}). \textsc{Zeltron} uses a standard explicit leapfrog discretization for particle and field equations and a numerical grid based on a Yee lattice. The RelIMM and RelSIM are implemented in the basic framework employed by \textsc{iPic3D} and \textsc{ECSIM}, i.e.\ a grid with colocated electromagnetic fields (see e.g.\ \citealt{markidis2010}) and the discretization discussed in the previous Sections for field and particle equations. {To solve the linear problem for the electric-field update, we employ a Jacobian-free Newton-Krylov iterative solver.}

\subsection{Beam Instabilities in 1D}

\begin{figure*}
\centering
\includegraphics[width=1\textwidth, trim={0mm 0mm 0mm 0mm}, clip]{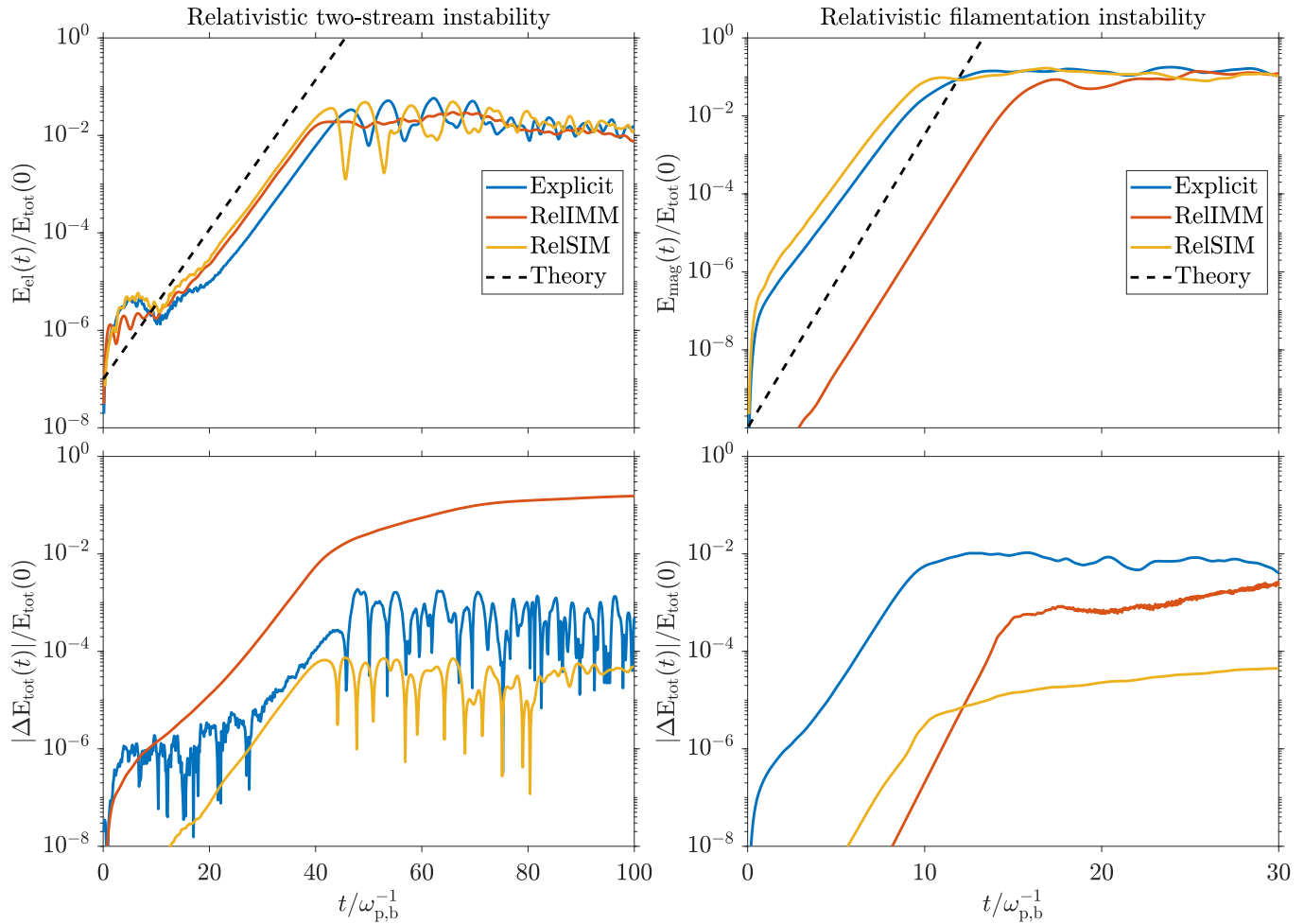}
\caption{Paradigmatic one-dimensional test cases for PIC methods, simulating the interaction of two counterpropagating electron-positron plasma beams with initial mean Lorentz factor $\gamma_0=2$. Left column: electric-energy evolution (top) and relative error on the total energy (bottom) for the electrostatic two-stream instability (where the beam velocity is along $x$). Right column: magnetic-energy evolution (top) and relative error on the total energy (bottom) for the electromagnetic filamentation instability (where the beam velocity is perpendicular to $x$). For both instabilities, the system's evolution follows the theoretical growth rate relatively well (before reaching a statistically similar nonlinear state) when employing an explicit-PIC method, the standard RelIMM from \citealt{Noguchi2007}, and the new RelSIM presented here. Errors in the energy are always much smaller for the new RelSIM with respect to the other two methods.}
\label{fig:beam}
\end{figure*}

As a first test, we consider the textbook case employed as a sanity check for every PIC code, i.e.\ a one-dimensional beam instability. In a 1D periodic domain $x\in[0,L]$ we initialize two counterpropagating neutral beams of electron-positron plasma (i.e.\ $m_i=m_e$). Particle velocities are drawn from a relativistic Maxwell-J\"{u}ttner distribution with mean Lorentz factor $\gamma_0=1/\sqrt{1-v_0^2/c^2}=2$ and a thermal spread $\Theta_0=kT_0/(m_e c^2)=0.001$. We consider both the simple electrostatic case in which the beam drift direction is along $x$ (i.e.\ a two-stream instability or TSI) and the electromagnetic case where the beams propagate perpendicularly to $x$ (i.e.\ a filamentation instability or FI). These instabilities have maximum growth rates $\Gamma^\mathrm{TSI}/\omega_\mathrm{p,b}=1/(2\gamma_0^{3/2})$ and $\Gamma^\mathrm{FI}/\omega_\mathrm{p,b}=(v_0/c)\sqrt{2/\gamma_0}$ (where $\omega_\mathrm{p,b}$ is the plasma frequency calculated with the density of a single beam; see e.g.\ \citealt{bret2010}) respectively. These two classical tests are useful to assess the basic properties of standard PIC schemes, and we simulate both with an explicit method as well as with the RelIMM and the RelSIM. In Fig.~\ref{fig:beam}, we show the results for both test cases.

For the TSI, the numerical domain is of size $L=32c/\omega_\mathrm{p,b}$ divided in 64 cells, with 156 particles per cell for each species (electrons and positrons). The Courant-Friedrichs-Lewy (CFL) ratio is kept such that $c\Delta t/\Delta x = 0.25$. The evolution of the electric energy is shown in the top-left panel of Fig.~\ref{fig:beam}, for the explicit-PIC case, the RelIMM, and the RelSIM. The reference theoretical growth rate is also shown for comparison. We observe that the electrostatic instability is captured well by all methods during the linear stage. The nonlinear stage shows (expected) differences between the methods, but an overall agreement in the saturation level of the electric energy. The bottom-left panel of the same Figure shows the evolution in time of the relative error on the total energy of the system, which should be conserved exactly in principle. We immediately notice that while the explicit approach and the new RelSIM keep energy errors well controlled, the RelIMM introduces much larger deviations, up to 10\% of the total energy.

For the electromagnetic FI, the domain is of size $L=12.8c/\omega_\mathrm{p,b}$ and we employ a grid with 256 cells and 20 particles per cell per species. The CFL ratio is such that $c\Delta t/\Delta x = 0.5$. The top-right panel of Fig.~\ref{fig:beam} shows the evolution of the magnetic energy, again for all methods, compared with the theoretical linear growth rate. We observe that all runs capture the linear stage relatively well, albeit with small deviations from theoretical expectations (potentially due to the low resolution employed). The saturation level during the nonlinear stage is again similar for all methods; however, it is interesting to notice that in this case the explicit method displays the largest energy errors, shown in the bottom-right panel of the same Figure. The RelIMM shows errors similar to the explicit method by the end of the run, while the RelSIM keeps energy errors roughly two orders of magnitude lower.

For the two very simple test cases considered, we conclude that all methods perform relatively well (as expected), at least in terms of capturing the linear stage of the instability. The RelSIM in particular distinguishes itself by introducing smaller errors on the total energy with respect to both the RelIMM and explicit methods, in all cases. This is an important property for numerical methods in general, but we will show in the next Sections that it can actually prove fundamental in physical cases of interest.

\subsection{Ion-electron Shock in 1D}

\begin{figure*}
\centering
\includegraphics[width=1\textwidth, trim={0mm 0mm 0mm 0mm}, clip]{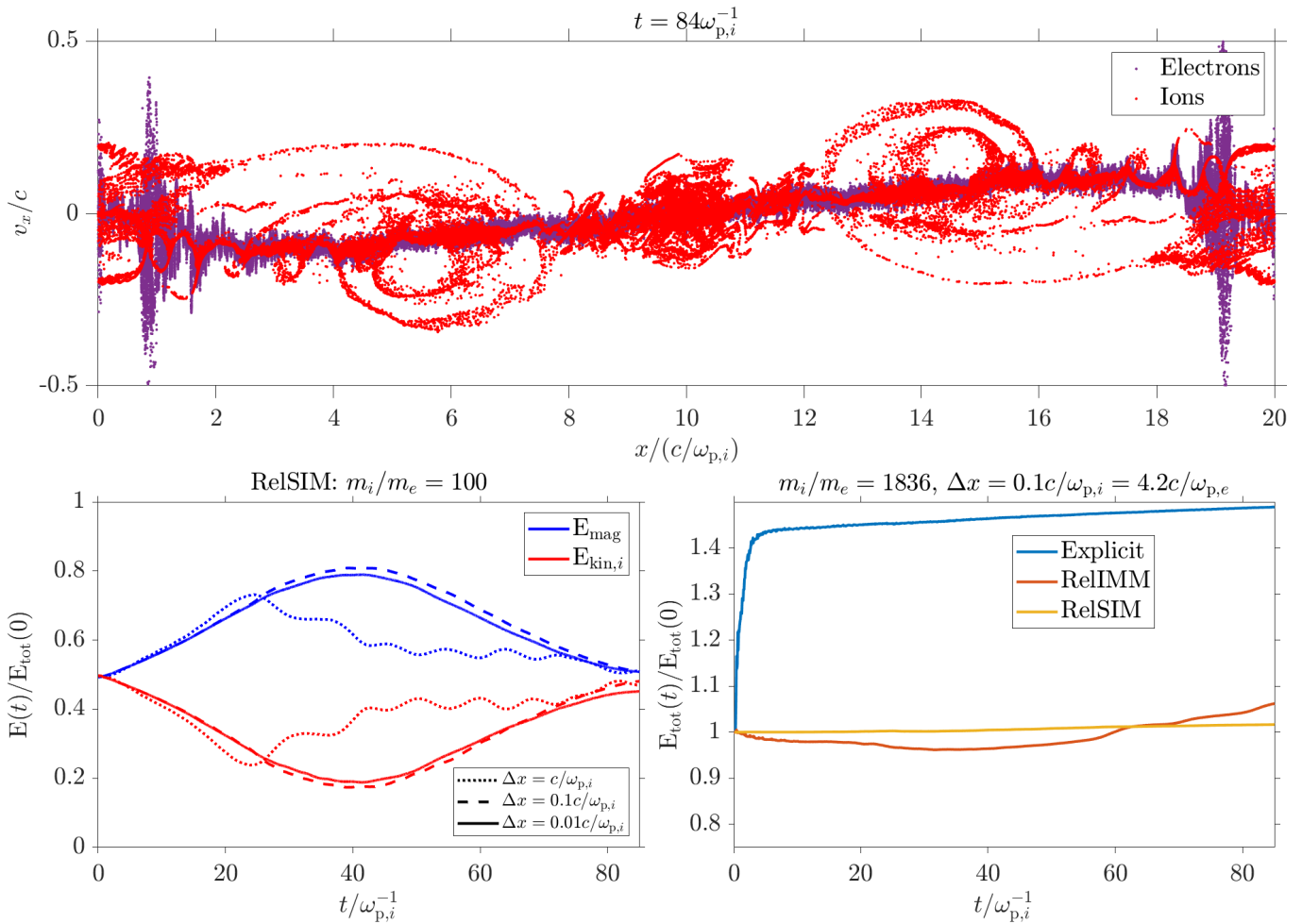}
\caption{One-dimensional ion-electron shock test. Two ion-electron beams propagating in a uniform out-of-plane magnetic field with opposite initial velocities collide along the $x$-direction and create outgoing shock waves. Top panel: Ions display a complex dynamics with multiple reflections at the shock fronts, while electrons mostly act as a thermal background. Bottom-left panel: The evolution in time of magnetic and ion energy shows that the new RelSIM produces converged results when the ion scales are well resolved, even if electron scales are underresolved. Bottom-right panel: The evolution of the total energy during the simulation shows that explicit PIC methods rapidly introduce large errors when electron scales are underresolved. Implicit methods remain stable, but the standard RelIMM performs much worse than the new RelSIM in terms of energy conservation.}
\label{fig:shock}
\end{figure*}

As a second test, we consider a one-dimensional shock problem in an ion-electron plasma with large mass ratio $m_i/m_e\gg 1$, which is relevant for e.g.\ particle acceleration at supernova remnants, plasma-expansion experiments, solar flares, and solar wind (\citealt{dieckmann2010,jones2011,park2013,liseykina2015,caprioli2018}). This test case is a representative example of a fully kinetic system where the global dynamics is almost entirely driven by the ions and occurs over ion-related length and time scales. Electrons mostly act as a background fluid, with little to no effect on the overall system evolution. Such a problem is challenging for explicit codes, which need to resolve all scales down to the electron skin depth and plasma frequency, whereas implicit codes in principle allow for underresolving electron scales.

To set up a simplified shock problem, we initialize two plasma beams of uniform density $n_0=n_{0,i}=n_{0,e}$ traveling in a uniform background magnetic field $\vecB_0=(0,0,B_0)$ such that the ion magnetization $\sigma_{0,i} = B_0^2/(4\pi n_0 m_i c^2) = 0.01$ everywhere. Ions and electrons in the beams have equal temperature $T_0$ such that the initial thermal spread $\Theta_{0,i} = \Theta_{0,e}/(m_i/m_e) = kT_0/(m_ic^2)=10^{-6}$. The beams are initialized in a domain $x\in[0,L]$ with initial mean velocity $\vecv_0=(\pm v_0,0,0)$ (with a $+$ sign if $x<L/2$ and a $-$ sign otherwise) where $v_0/c=0.1$. The domain size here is $L=20\tilde{\rho}_{\mathrm{C},i}$, where $\tilde{\rho}_{\mathrm{C},i}\equiv mc v_0/(qB_0)$ (note that with the chosen $\sigma_{0,i}=0.01$, $\tilde{\rho}_{\mathrm{C},i}=c/\ompi$). Finally, the initial electric field is set equal to $-\vecv_0\btimes\vecB_0$. With this setup, two shock waves are created at $t=0$ at the domain center; both shocks then travel toward the boundaries with speed $\sim v_0$. Although the initial state is nonperiodic in nature, we employ periodic boundary conditions for simplicity, assuming that spurious boundary effects do not drastically influence the solution until the shocks reach $x=0$ or $x=L$ around $t\simeq 100\ompi^{-1}$. For this reason, we halt the simulation at $t= 85\ompi^{-1}$, before boundary effects come into play.

We simulate our simple shock setup with an explicit method as well as with the RelIMM and the new RelSIM. In the top panel of Fig.~\ref{fig:shock}, we show a representative snapshot of both ions and electrons in the $x-v_x$ phase space at $t=84\ompi^{-1}$. The shock fronts are visible at $x\simeq L/2\pm 9c/\ompi$. The ions (in red) show signatures of multiple reflections in the shock downstream, and a generally complex distribution in phase space as a result of the shock propagation. Electrons (in purple), instead, are predominantly behaving as a thermal background.

We first test the convergence of the RelSIM with respect to the numerical resolution. Fixing $c\Delta t/\Delta x = 0.7$, we employ a mass ratio $m_i/m_e=100$ and vary the grid spacing $\Delta x/(c/\ompi) = 1, 0.1, 0.01$. In terms of electron scales, this corresponds to $\Delta x/(c/\ompe) = 10, 1, 0.1$, i.e.\ the electron skin depth goes from dramatically underresolved to relatively well resolved. In all runs we initialize 100 particles per cell per species. In the bottom-left panel of Fig.~\ref{fig:shock}, we show the results of this test in terms of the evolution of total magnetic and ion energy. We observe that the results converge when the ion scales are well resolved by the computational grid, i.e.\ further increasing the resolution beyond $\Delta x/(c/\ompi) = 0.1$ does not dramatically alter the evolution of the system. A substantial difference arises when ion scales are only marginally resolved ($\Delta x/(c/\ompi) = 1$) which is unsurprising, given that the characteristic length scales of the problem are not well captured.

As a second test, we consider the exact same initial conditions but with a realistic mass ratio $m_i/m_e=1836$. In this case, electron and ion scales are even more separated and our reference resolutions $\Delta x/(c/\ompi) = 1, 0.1, 0.01$ are such that $\Delta x/(c/\ompe) \simeq 42, 4.2, 0.42$ in terms of electron scales. Moreover, from our choice $c\Delta t/\Delta x = 0.7$ it follows that $\Delta t/\ompe^{-1} \simeq 29, 2.9, 0.29$, i.e.\ both electron length and time scales are  largely underresolved in two out of three runs. The results for the case $\Delta x/(c/\ompi) = 0.1$ (i.e.\ $\Delta x/(c/\ompe) \simeq 4.2$) are shown in the bottom-right panel of Fig.~\ref{fig:shock} in terms of the evolution of the total energy of the system, which should be exactly conserved, for the explicit run and for the RelIMM and RelSIM cases. For the explicit method, we observe that errors in the total energy rapidly increase right from the start of the simulation. We find that this energy error precisely corresponds to an unphysical increase in electron energy (not shown) that occurs when electron scales are underresolved in the explicit run. The RelIMM and RelSIM runs behave much better, introducing much smaller errors; in particular, the RelSIM displays the smallest errors out of the three methods. The same qualitative conclusion applies to all runs conducted here, including those not shown in Fig.~\ref{fig:shock}.

These results show that the new RelSIM is superior to the standard explicit PIC when underresolving electron scales. This is rather unsurprising, given that the implicit approach eliminates stability constrains affecting explicit methods; however, the RelSIM also produces better results than those obtained with the original RelIMM, by introducing smaller energy errors in all cases. An additional interesting feature is that the RelSIM retains stability even in demanding cases in which ion scales are marginally resolved or underresolved. As shown in Fig.~\ref{fig:shock}, when ion scales are not accurately captured the dynamics of the system at those scales is approximated, but not completely lost;  underresolving ion scales does not result in a loss of stability of the method, i.e.\ the underresolved simulation is not disrupted by numerical artifacts growing unboundedly.

\subsection{Relativistic Reconnection in 2D}

\begin{figure*}
\centering
\includegraphics[width=1\textwidth, trim={0mm 55mm 0mm 55mm}, clip]{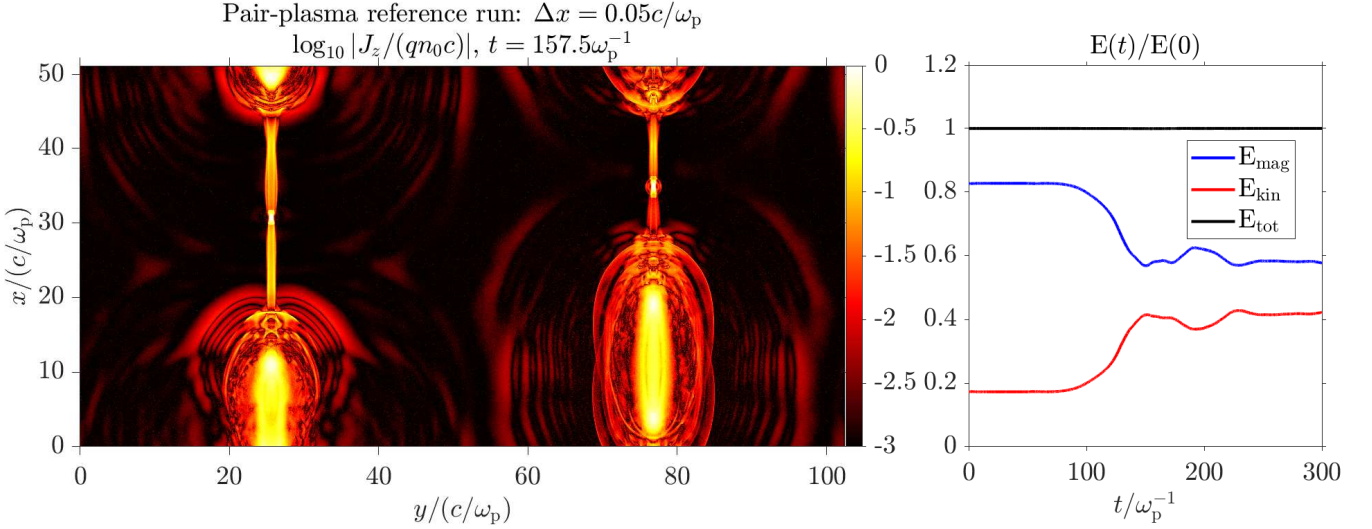}
\caption{Representative pair-plasma reconnection simulation at high numerical resolution. Left: Spatial distribution of the out-of-plane current density during the nonlinear stage of the evolution, showing the typical ``plasmoid'' structures originating from the fragmentation of the initial current sheets. Right: Typical evolution of the system's energetics, where magnetic energy (in blue) is converted into kinetic energy (in red) while the total energy (in black) is conserved.}
\label{fig:recon_ref}
\end{figure*}

In this Section we consider a two-dimensional setup for magnetic reconnection as a test for PIC algorithms in multiple dimensions. Reconnection is a ubiquitous process in the Universe, conjectured to play a major role in many high-energy astrophysical environments as well as in solar, heliospheric, and fusion plasmas (see e.g.\ \citealt{hoshinolyubarsky2012} for a review on relativistic reconnection). We initialize two relativistic Harris sheets (e.g.\ \citealt{harris1962,melzani2013}) in a double-periodic domain $(x,y)\in [L_x\times L_y]$ where an upstream, magnetized ion-electron thermal plasma flows into a region of magnetic-polarity inversion and experiences acceleration as reconnection dissipates magnetic energy. To set up the initial Harris equilibrium we first impose the upstream conditions in terms of the background ion thermal spread $\Theta_{0,i} = kT_0/{m_i c^2}$ and ion magnetization $\sigma_{0,i} = B_0^2/(4\pi n_0 m_i c^2)$ where we have assumed that $n_0=n_{0,i}=n_{0,e}$ and $T_0=T_{0,i}=T_{0,e}$. The ion temperature and magnetization are free parameters, while the background electron thermal spread follows from $\Theta_{0,e} = \Theta_{0,i}(m_i/m_e)$. Then, the plasma conditions inside the current sheets can be calculated from the initial magnetic-field profile,
\be
B_x(y) = 
\begin{cases}
-B_0\tanh\left(\frac{y-L_y/4}{\delta}\right) & \mbox{if} \quad y<L_y/2 \\
B_0\tanh\left(\frac{y-3L_y/4}{\delta}\right) & \mbox{if} \quad y>L_y/2
\end{cases},
\ee
where $\delta$ is the current-sheet half-thickness (a free parameter of the setup). By imposing  $c\grad\btimes\vecB=4\pi\vecJ$ and pressure balance across a current sheet, we can find the plasma drift velocity $\vecv_\mathrm{CS}=(0,0,\pm v_\mathrm{CS}$) (equal and opposite for the two species) and temperature at the current-sheet center (e.g.\ \citealt{melzani2013}),
\be
\frac{v_{\mathrm{CS}}\Gamma_\mathrm{CS}}{c} = \frac{B_0}{8\pi q\alpha n_0\delta},
\ee
\be
\Theta_{\mathrm{CS},i} = \Theta_{\mathrm{CS},e}/(m_i/m_e) = \frac{B_0^2\Gamma_\mathrm{CS}}{16\pi \alpha n_{0} m_i c^2} = \frac{\sigma_{i,0}\Gamma_\mathrm{CS}}{2},
\ee
where $q=|q_i|=|q_e|$ and $\alpha$ is the ratio of plasma density between current-sheet center and upstream. The drift motion determines a Lorentz factor $\Gamma_\mathrm{CS} = 1/\sqrt{1-v_\mathrm{CS}^2/c^2}$ of the drifting plasma inside the current sheet. The overdensity ratio $\alpha$ is a free parameter like $\delta$, but note that $\alpha$ and $\delta$ must be chosen\footnote{Choosing $\alpha$ and $\delta$ such that $v_\mathrm{CS}>c$ is equivalent to choosing parameters for which the Harris equilibrium cannot be satisfied.} such that $v_\mathrm{CS}<c$.

As a first test, we consider the simple pair-plasma case, $m_i=m_e$, and impose an upstream magnetization $\sigma_0=\sigma_{0,i}=\sigma_{0,e}=10$ (calculated with both species) and temperature $\Theta_0=\Theta_{0,i}=\Theta_{0,e} = 0.01$. We consider a domain of size $L_x=L_y/2 = 51.2 c/\omp$, where $\omp$ includes the density of both species combined. The current-sheet half-thickness $\delta/(c/\omp)=1$ and the overdensity ratio $\alpha=5$. We perform a set of simulations where we progressively decrease the grid spacing $\Delta x/(c/\omp) = 4,2,1,0.5$ keeping $c\Delta t/\Delta x = 1/\sqrt{2}$ fixed. In all cases we initialize 64 particles per cell per species, and we do not perturb the initial equilibrium, such that the onset of the tearing instability leading to reconnection is only determined by numerical noise. We employ a high-resolution simulation with $\Delta x/(c/\omp) = 0.05$ as a reference result. A representative snapshot of the reference solution when reconnection is fully developed is shown in Fig.~\ref{fig:recon_ref}: the out-of-plane current-density  distribution during the nonlinear stage of the simulation (left panel) features the typical ``plasmoids'' created by the fragmentation of the initial current sheets. The evolution of the system's energetics (right panel) is such that magnetic energy is depleted in favor of kinetic energy, before reaching a statistical steady state. Our system size is relatively small, such that the reconnection process is halted within a few hundred plasma periods.

\begin{figure*}
\centering
\includegraphics[width=1\textwidth, trim={0mm 0mm 0mm 0mm}, clip]{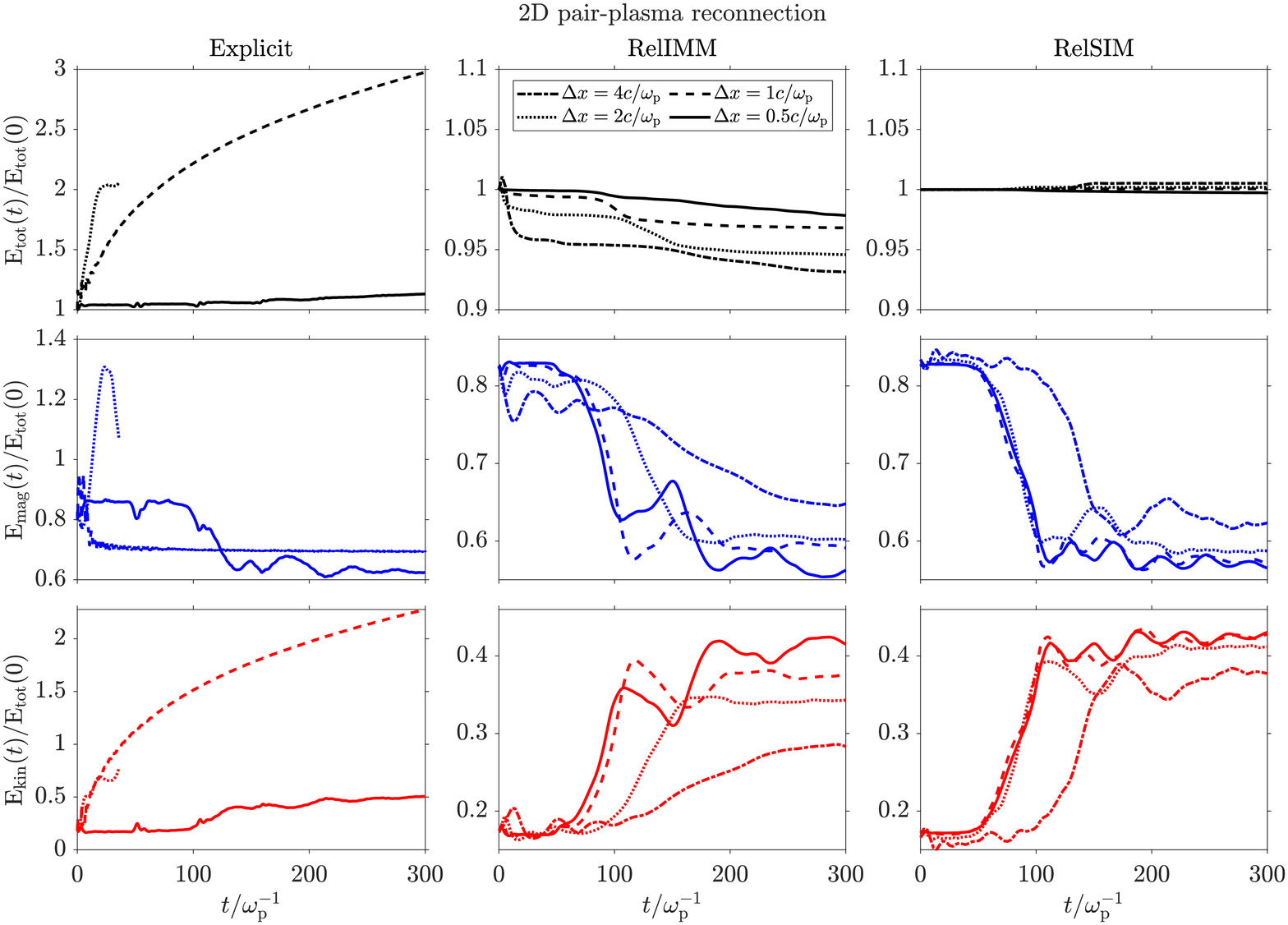}
\caption{Simulations of pair-plasma reconnection at different resolutions with an explicit PIC method (left column), the RelIMM (middle column), and the RelSIM (right column). For an array of grid spacings $\Delta x/(c/\omp)=4,2,1,0.5$ with 64 particles per cell per species (electrons and positrons), we show the evolution of the total (top row), magnetic (middle row), and kinetic (bottom row) energy. Note that when $\Delta x/(c/\omp)=4$ the explicit simulation fails after only a few time steps, hence the corresponding lines are not shown in the left column.}
\label{fig:recon_pairs}
\end{figure*}

Fig.~\ref{fig:recon_pairs} shows the results, for the resolutions indicated above, of the pair-plasma simulation run with the explicit method (left column), the RelIMM (middle column), and the RelSIM (right column) in terms of the total (top row), magnetic (middle row), and kinetic (bottom row) energy over time. For the explicit method, we observe how numerical errors rapidly destroy the solution when $\Delta x> c/\omp$ (note that the case $\Delta x=4c/\omp$ immediately fails after a few time steps and is therefore not shown). Both implicit methods, instead, remain stable for all resolutions considered. As the number of grid points decreases the solution becomes less accurate, and particles and fields exchange less and less energy. It is however interesting to note that the RelSIM already provides relatively well-converged results for $\Delta x\le 2c/\omp$, at least in terms of the rate of depletion of magnetic energy (and the corresponding increase in kinetic energy). Conversely, the RelIMM still shows larger differences in the solutions between $\Delta x\le c/\omp$ and $\Delta x\le 0.5c/\omp$. The evolution of the total energy for the two implicit methods appears tightly linked with how well those methods converge: the RelSIM systematically shows much lower energy errors, and converges faster, than the RelIMM, suggesting that energy conservation is of primary importance to produce qualitatively accurate results even at low resolutions.

As a second test case, we consider an ion-electron plasma with realistic mass ratio $m_i/m_e=1836$. We initialize the system similarly to the pair-plasma case, but with important differences. The upstream ion magnetization and temperature are $\sigma_{0,i}=10$ and $\Theta_{0,i} = 0.01$, and by choosing $T_{0,e}=T_{0,i}$ this implies $\Theta_{0,e} = 18.36$, i.e.\ upstream electrons are relativistically hot (with mean Lorentz factor $\gamma_{0,e}\simeq 55$). The domain size is $L_x=L_y/2 = 51.2 c/\ompi$, and we choose $\delta/(c/\ompi)=1$ and $\alpha=5$. Because $\gamma_{0,e}\gg1$, this setup is representative of the so-called ``semirelativistic'' reconnection regime, relevant for e.g.\ accreting black-hole coronae and blazar jets {(e.g.\ \citealt{rowan2017,ball2018,werner2018,kilian2020})}. With respect to the pair-plasma case, ion and electron spatiotemporal scales are now separated, but less than they would be in a completely nonrelativistic scenario: the relativistic electron skin depth is indeed $c/\ompe^\mathrm{r}=c\sqrt{\gamma_{0,e}}/\ompe\simeq0.17c/\ompi$, i.e.\ a factor $\sqrt{\gamma_{0,e}}$ larger than the corresponding nonrelativistic counterpart. Since relativistic explicit PIC codes must resolve $c/\ompe^\mathrm{r}$ on the numerical grid, the presence of relativistic electrons helps relaxing the stability criterion for explicit simulations. In our test, we employ numerical resolutions $\Delta x/(c/\ompi) = 4,2,1,0.5$ corresponding to $\Delta x/(c/\ompe^\mathrm{r}) \simeq 23.1, 11.6, 5.8, 2.9$; we also keep $c\Delta t/\Delta x = 1/\sqrt{2}$ fixed. As a result, in all cases the electron spatial and temporal scales are underresolved. 

\begin{figure*}
\centering
\includegraphics[width=1\textwidth, trim={0mm 0mm 0mm 0mm}, clip]{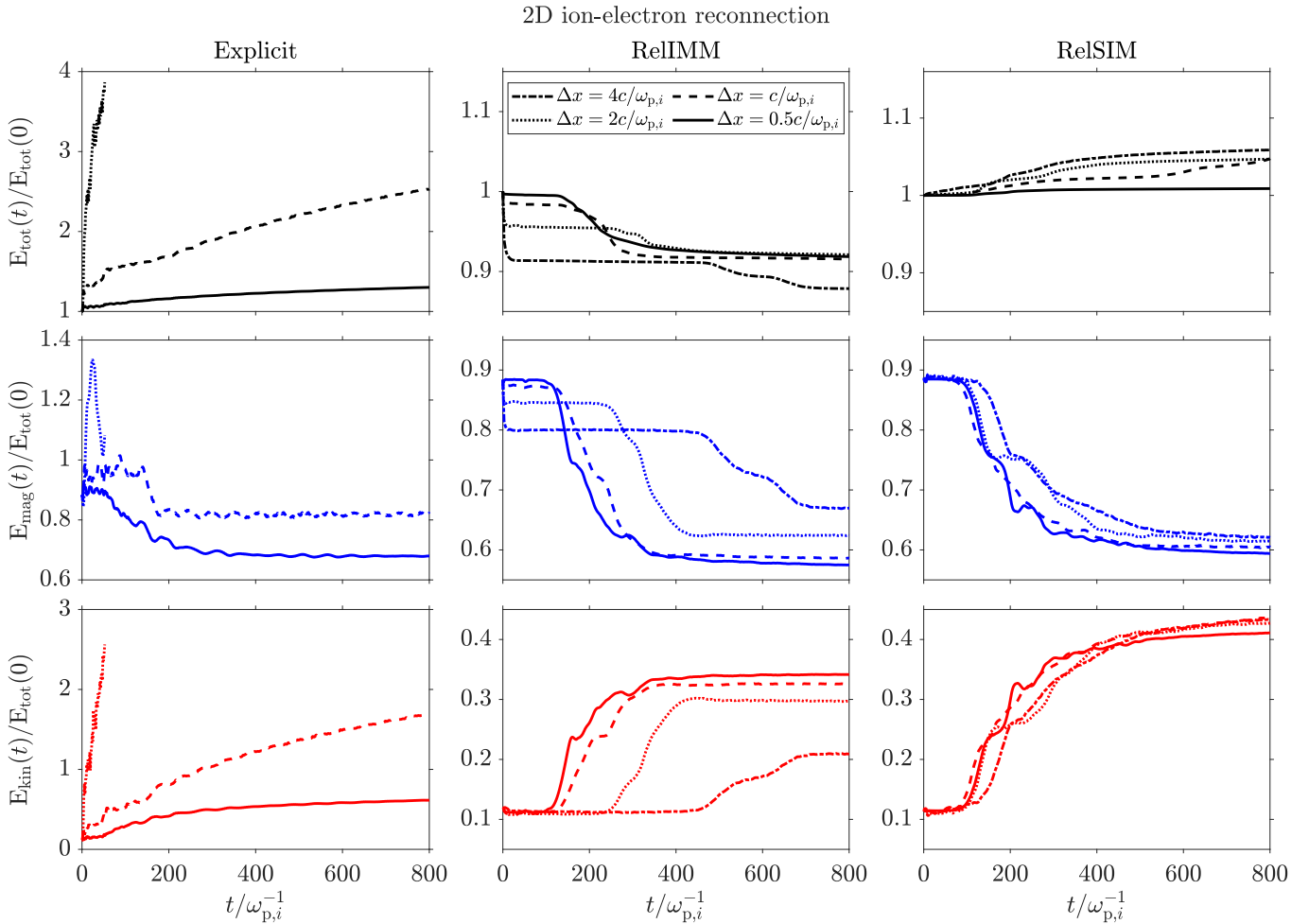}
\caption{As in Fig.~\ref{fig:recon_pairs} but for an ion-electron reconnection setup with realistic mass ratio $m_i/m_e=1836$.}
\label{fig:recon_ionel}
\end{figure*}

Fig.~\ref{fig:recon_ionel} shows the same quantities as Fig.~\ref{fig:recon_pairs}, now for the ion-electron case: the evolution of total, magnetic, and total kinetic energy (i.e.\ of ions and electrons combined) for our array of simulations using the explicit method, the RelIMM, and the RelSIM. The results are very similar to the pair-plasma case: while the explicit runs either quickly fail or display large numerical errors, the implicit runs remain stable even when both electron and ion scales are dramatically underresolved. The RelSIM again performs systematically better than the RelIMM, introducing smaller numerical errors and converging much faster to the expected behavior (i.e.\ the dissipation of magnetic energy corresponding to an increase of kinetic energy).

\begin{figure}
\centering
\includegraphics[width=1\columnwidth, trim={0mm 0mm 0mm 0mm}, clip]{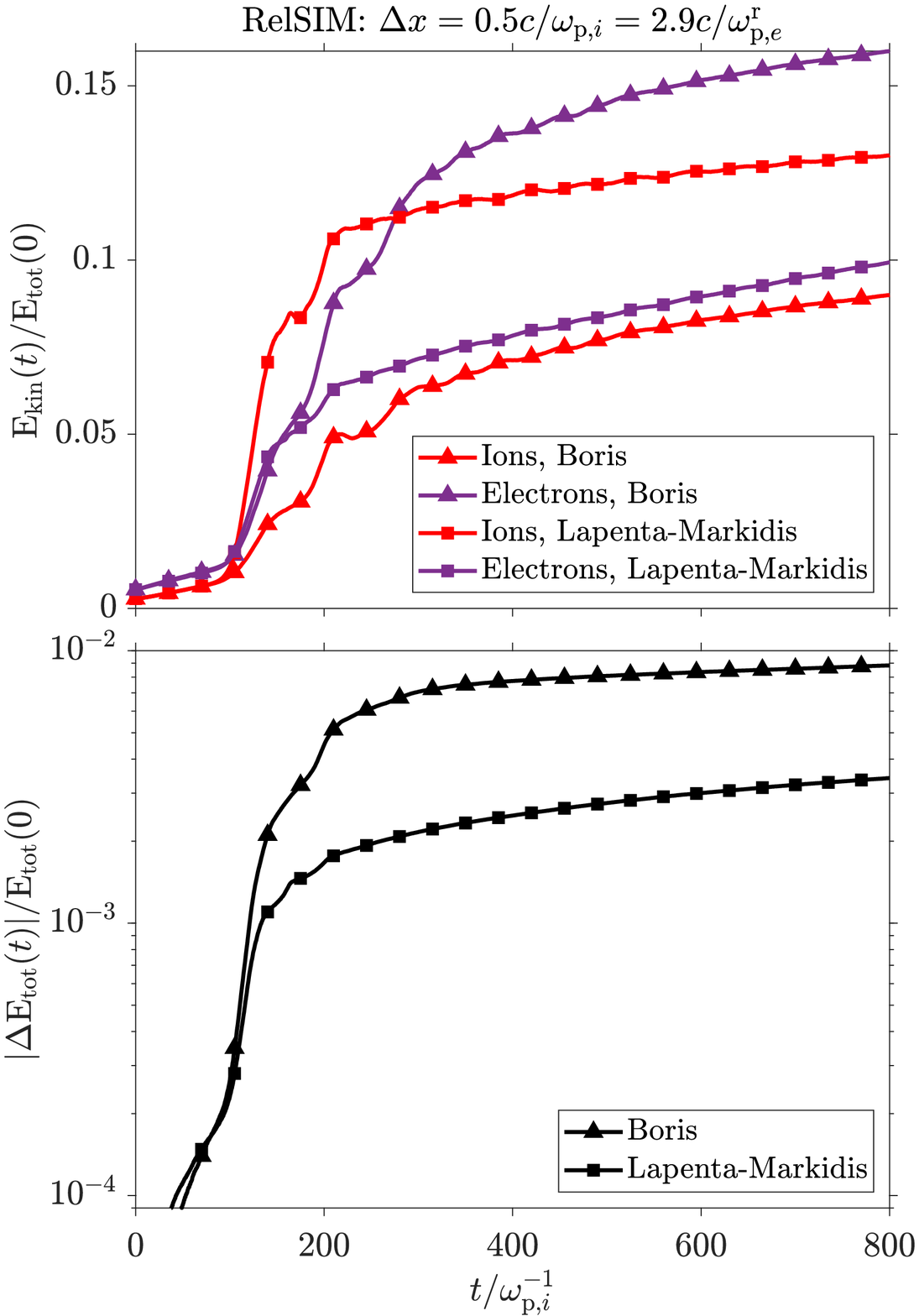}
\caption{Comparison of different particle pushers applied to representative ion-electron reconnection problem with resolution $\Delta x=0.5c/\ompi$. Top panel: Evolution in time of the ion (in red) and electron (in purple) kinetic energy for two runs employing the standard Boris pusher and the Lapenta-Markidis pusher (triangle and square markers respectively). Bottom panel: The evolution of the relative error in the total energy for the same runs.}
\label{fig:recon_LMvsBoris}
\end{figure}

As a last experiment, we consider the effect of different particle pushers on the performance of the RelSIM in the ion-electron case. While the global evolution of the reconnection layers is similar between the $m_i=m_e$ and $m_i\gg m_e$ cases, the individual species behave differently in the latter scenario and in particular they receive different amounts of energy from the reconnection process (e.g.\ \citealt{werner2018}). In Fig.~\ref{fig:recon_LMvsBoris} (top panel), we plot the evolution in time of the ion (in red) and electron (in purple) kinetic energy during the reconnection simulation run with the RelSIM and with resolution $\Delta x=0.5c/\ompi$. We distinguish between the evolution produced by the Boris and the Lapenta-Markidis pushers. Interestingly, we observe a completely opposite behavior in the two cases: with the Boris pusher, electrons gain more energy than ions, while the reverse occurs with the Lapenta-Markidis pusher. The latter behavior (ions gaining larger amounts of energy), for our choice of relatively low ion magnetization $\sigma_{0,i}=10$, corresponds to theoretical expectations and earlier numerical experiments (for much larger magnetizations the two species gain approximately the same amount of energy; see \citealt{werner2018}). Hence, we conclude that the Lapenta-Markidis pusher produces a more accurate result for this specific physical case. It is also instructive to measure energy conservation in this run, which we show in the bottom panel of Fig.~\ref{fig:recon_LMvsBoris}. The evolution of the energy error in the two simulations is such that the Lapenta-Markidis pusher produces smaller energy errors at all times, resulting in a $\sim 3\%$ energy deviation at the end of the run, while the Boris pusher reaches errors around 3 times larger while also inverting the behavior of electron and ion energy gain. This result is not particularly surprising, considering that the Lapenta-Markidis pusher intrinsically possess superior energy-conservation properties, as discussed in Section~\ref{sec:RelIMM}.

Our experiments for the simple two-dimensional reconnection setup considered here leads us to conclude that the new RelSIM performs much better than both the explicit method and the original RelIMM, producing physically meaning results even at very low resolutions where ion and electron time and length scales are dramatically underresolved.

\section{Discussion and summary}
\label{sec:conclusions}
We have presented a novel Relativistic Semi-Implicit Method (RelSIM) for fully kinetic simulations of astrophysical plasmas. Implicit PIC methods in general possess superior stability and energy-conservation properties with respect to standard explicit methods, but an implicit \emph{relativistic} PIC method suitable for production runs is currently missing from the panorama of available approaches. We propose the RelSIM, currently implemented in the framework of the \textsc{ECsim} code (\citealt{lapenta2017ecsimcode}), as a production-ready tool for large-scale PIC simulations. The work presented in this paper can be summarized as follows:
\begin{itemize}
    \item We have reviewed the Relativistic Implicit-Moment Method (RelIMM), originally presented by \cite{Noguchi2007}, generalizing it to be compatible with different particle pushers available in literature. In doing so, we have also presented for the first time an explicit solution for the Lapenta-Markidis relativistic pusher (\citealt{LapentaMarkidis2011}), which provides better energy conservation with respect to other standard approaches such as the Boris pusher.
    \item We have constructed the new RelSIM method as a relativistic extension of the Energy-Conserving Semi-Implicit method (ECSIM) first presented in \cite{lapenta2017ecsimcode}. The new method can also be employed with several particle pushers available in literature. To derive the RelSIM, we have introduced one single approximation in the discrete Vlasov-Maxwell system, in contrast with several heavy approximations on which the RelIMM is based. The RelSIM is also free of nonlinear iterations, only requiring a linear solver on the field quantities stored on the grid.
    \item We have thoroughly tested the RelSIM in a number of idealized setups in one and two spatial dimensions, comparing its performance to that of the RelIMM and of a standard explicit-leapfrog PIC method (implemented in the state-of-the-art code \textsc{Zeltron}). For this purpose, we have employed idealized setups for one-dimensional beam instabilities in pair plasmas, a one-dimensional ion-electron shock, and a two-dimensional reconnection setup for pair plasmas and for ion-electron plasmas.
\end{itemize}

In all our experiments, the RelSIM performs distinctively better than both a standard explicit method and the original RelIMM. We have quantified this performance in terms of i) errors in the total energy of the system; ii) stability and convergence of the method when relevant plasma scales (e.g.\ ion and electron skin depth and plasma period) are dramatically underresolved; and iii) behavior with different particle pushers. In our tests, we found that the RelSIM produces much smaller energy errors than the other methods, and that compared to explicit methods it retains stability even when plasma scales are underresolved by orders of magnitude (as expected). In these underresolved cases, explicit PIC methods rapidly produce unphysical results, while the RelSIM simply approximates the solution capturing the physics correctly up to the resolved scales. In our two-dimensional relativistic reconnection simulation of an ion-electron plasma with realistic mass ratio, we also found that using the Lapenta-Markidis integrator produces more physically realistic results than the standard Boris integrator.

First-principles simulations in the collisionless regime are an extremely powerful tool to study relativistic plasmas, but are often limited by numerical constraints imposed by standard explicit methods. We showed here that implicit PIC methods such as the new RelSIM do not suffer from these limitations; while more computationally intensive than explicit PIC, the RelSIM compensates by allowing for lower resolutions when it is not necessary to resolve the smallest scales in a certain system. We provided an example by simulating an ion-electron shock case in one dimension, where the scales of the problem are those of ions, and electrons mostly provide a neutralizing background. In this case, it is interesting to retain the electron physics since high-energy electrons could in principle interact with ion-scale structures and experience acceleration; but the problem does not intrinsically require the full modeling of electron scales, which explicit methods are bound to resolve. We showed that the RelSIM can indeed completely neglect electron-scale physics while retaining stability, allowing for cheap ion-scale simulations that also include kinetic electrons, in contrast with e.g.\ hybrid methods.

In multidimensional simulations, the gain factor of implicit methods {is even larger}, because computational time can be saved by reducing the grid resolution along each spatial dimension. We provided an example with a paradigmatic two-dimensional reconnection problem, where the RelSIM retains stability even with a coarse grid resolution that underresolves the largest scales. In such a scenario, it is {legitimate} to question whether employing a poor numerical resolution makes sense at all, since in doing so the reconnection physics may be lost. In such a case, we envision the application of our method in combination with a nonuniform grid (e.g.\ \citealt{chaconchen2016}; Croonen et al. 2023, in prep.) that concentrates resolution in the reconnection region, while dramatically underresolving the upstream-plasma scales (where plasma simply flows uniformly toward the current sheets). In this way, resolved reconnection physics could be retained while also speeding up calculations in the upstream without loss of stability.

While the RelSIM is in principle ready for production runs, ample ground is available for improvements and future developments:
\begin{itemize}
    \item Differently from its nonrelativistic counterpart, the RelSIM does not conserve energy to machine precision, due to intrinsic nonlinearities in the field equations. By removing these nonlinearities we obtain a simpler, linear system, but we also introduce (small) energy errors. It is in principle possible to retain exact energy conservation by discarding our approximation and iterating on the field-particle equations up to exact nonlinear convergence; alternatively, a fixed amount of iterations could also help in improving energy conservation without reaching exact accuracy (see e.g.\ \citealt{angus2023}).
    \item While in our tests we have employed the strategy by \citealt{chentoth2019} to ensure that Gauss's law for $\vecE$ is satisfied, the algorithm does not by default conserve charge. This is because by construction we cannot adopt charge-conserving deposition schemes (e.g.\ \citealt{villasenorbuneman1992,esirkepov2001}) in our implicit method\footnote{The same applies for the IMM and ECSIM family of methods.}. As was shown for the nonrelativistic ECSIM, charge conservation can be imposed exactly in several ways (e.g.\ \citealt{chentoth2019,campospintopages2022}), or approximately via divergence-cleaning schemes (e.g.\ \citealt{marder1987}); in the future, we will explore different strategies to impose charge conservation in the RelSIM optimally.
    \item To construct our method, we have also staggered particle positions and velocities in time, such that no iteration is needed to advance the particles. Similarly, we have also decentered $\vecB$ in the particle momentum equation, such that the magnetic field needed to advance the particles is immediately available. These choices may influence the behavior of the method in specific cases, e.g.\ by modifying the $\vecE\btimes\vecB$-motion of particles in electromagnetic fields. We will consider these numerical issues and possible improvements in future work (see also \citealt{angus2023} and references therein).
    \item In our first implementation, we have not considered the possibility of subcycling on the particle update or to employ smoothing to combat numerical noise in the solution. Both operations could be readily added to the RelSIM in exactly the same fashion that is employed for the nonrelativistic ECSIM (e.g.\ \citealt{lapenta2023}). We leave the exploration of these possibilities for future work. 
\end{itemize}

In summary, even in our first implementation, the new RelSIM provides a reliable, production-ready alternative to standard explicit PIC methods for astrophysical plasma simulations. We specifically target scenarios where large scale separation exists between different plasma species, or where the physics of interest only occurs in localized regions, or where the time and length scales involved become prohibitive for explicit approaches. Our method could also be combined with existing hybrid approaches for multiscale simulations (e.g.\ \citealt{toth2016,markidis2018,bacchini2020}), to further extend its reach to even larger-scale systems. In such situations, the RelSIM could provide dramatic speedup, helping to probe astrophysical regimes so far inaccessible with state-of-the-art codes.

\section*{Acknowledgements}
The author would like to thank Sasha Philippov, Anatoly Spitkovsky, Jean-Luc Vay, Stefano Markidis, Giuseppe Arr\`{o}, and Giovanni Lapenta for useful discussions and suggestions throughout the development of this work.
F.B.\ acknowledges support from the FED-tWIN programme (profile Prf-2020-004, project ``ENERGY'') issued by BELSPO.
The computational resources and services used in this work were provided by the VSC (Flemish Supercomputer Center), funded by the Research Foundation Flanders (FWO) and the Flemish Government – department EWI.
This work was performed in part at Aspen Center for Physics, which is supported by National Science Foundation grant PHY-1607611.

\appendix

\section{Explicit solution for the Lapenta-Markidis momentum update}
\label{app:LMsolution}

The Lapenta-Markidis particle mover (\citealt{LapentaMarkidis2011}) possesses superior energy-conservation properties with respect to other standard relativistic pushers. However, contrary to other popular movers, no explicit solution of the momentum equation employed in this approach has been presented in literature. We report such a solution here for the first time. Recall, for each particle $p$, that for the Lapenta-Markidis definition $\vecvbar_p =\vecubar_p/\bgam_p$, $\bgam_p=(\gamma_p^{n+1}+\gamma_p^n)/2$, and dotting eq.~\eqref{eq:mom_discrete} with $\vecvbar_p$ and rearranging terms gives
\be
c^2\bgam_p(\bgam_p-\gamma_p^n) = \frac{q_p\Delta t}{2 m_p}\vecE_p^{n+\theta}\bcdot\vecubar_p,
\ee
where $\vecubar_p = (\vecu_p^{n+1}+\vecu_p^n)/2$. Now, from eq.~\eqref{eq:mom_discrete} we can write an explicit expression for $\vecubar_p$ in terms of $\bgam_p$, following exactly the same procedure that allowed us to write eq.~\eqref{eq:vbar_imm}: with the shorthand notation $\bb{\beta}_p=q_p\Delta t\vecB_p^n/(2m_pc)$, $\bb{\epsilon}_p=q_p\Delta t\vecE_p^{n+\theta}/(2m_p)$, $\vecu'_p = \vecu_p^n+\veceps_p$,
\be
\vecubar_p = \frac{\vecu'_p+(\vecu'_p\bcdot\vecbeta_p)\vecbeta_p/\bgam_p^2+\vecu'_p\btimes\vecbeta_p/\bgam_p}{1+\beta_p^2/\bgam_p^2}.
\label{eq:ubarlapentamarkidis}
\ee
Combining these two equations provides a fourth-order polynomial,
\be
 -\bgam_p^4 + \gamma_p^n\bgam_p^3 + \xi\bgam_p^2 + \eta\bgam_p + \zeta = 0,
\ee
where the coefficients of the polynomial are $\xi=\vecu_p'\bcdot\veceps_p/c^2-\beta_p^2$, $\eta=(\vecu_p'\btimes\vecbeta_p)\bcdot\veceps_p/c^2+\vecbeta_p^2\gamma_p^n$, and $\zeta=(\vecu_p'\bcdot\vecbeta_p)(\vecbeta_p\bcdot\veceps_p/c^2)$. Solving for $\bgam$ can be done with any preferred method, and we find that a direct solution provides the fastest result: root analysis shows that $\bgam\ge1$ only for
\be
\bgam_p = \frac{\gamma_p^n}{4} + \frac{1}{2}\sqrt{2P+\frac{Q}{4\sqrt{P+R}}-R} + \frac{1}{2}\sqrt{P+R},
\ee
where
\begin{equation*}
 P = \frac{2}{3}\xi+\frac{(\gamma_p^n)^2}{4}, \qquad Q = 4\xi\gamma^n+8\eta+(\gamma_p^n)^3,
\end{equation*}
\begin{equation*}
R = \frac{S}{3T}+\frac{T}{3}, \qquad S = \xi^2-3\eta\gamma_p^n-12\zeta,
\end{equation*}
\begin{equation*}
T = \sqrt[3]{\frac{U+\sqrt{U^2-4S^3}}{2}}, \qquad U = -2\xi^3+9\xi \eta\gamma_p^n-72\xi\zeta+27\eta^2-27\zeta(\gamma_p^n)^2.
\end{equation*}
Pathological cases for this solution exist but are straightforward to handle, e.g.\ when $\veceps_p=\bb{0}$ the solution reduces to $\bgam_p=\gamma_p^n$. Once $\bgam_p$ is known, the new particle 4-velocity can be calculated from eq.~\eqref{eq:ubarlapentamarkidis} via extrapolation, $\vecu_p^{n+1}=2\vecubar_p-\vecu_p^n$.


\bibliographystyle{aasjournal}
\end{document}